\pdfoutput=1 
\documentclass[entropy,aricle,accept,pdftex,oneauthor]{Definitions/mdpi} 

\firstpage{1} 
\pubvolume{25}
\issuenum{1}
\articlenumber{648}
\pubyear{2023}
\copyrightyear{2023}
\externaleditor {{Academic Editor: Alessandro Giuliani}} 
\datereceived{23 February 2023} 
\daterevised{6 April 2023} 
\dateaccepted{7 April 2023} 
\datepublished{12 April 2023} 
\hreflink{https://doi.org/10.3390/e25040648} 

\usepackage{amsmath}
\usepackage{amssymb}
\usepackage{amsthm}
\usepackage{mathtools}

\newtheorem{defn}{Definition}

\usepackage{xcolor}
\definecolor{Crimson}{rgb}{0.6471, 0.1098, 0.1882}
\definecolor{Aqua}{rgb}{0.098, 0.54117, 0.68627}
\definecolor{cbGreen}{HTML}{4DAC26}
\definecolor{cbRed}{HTML}{D01C8B}
\usepackage{graphicx}
\graphicspath{{images/figs}}
\usepackage{placeins}

\usepackage{tikz}
\usetikzlibrary{arrows.meta}
\usepackage{pgfplots}
\usepgfplotslibrary{fillbetween}
\usetikzlibrary{intersections}
\pgfplotsset{compat=1.14}

\pgfdeclarelayer{bg}
\pgfsetlayers{bg,main}
\usetikzlibrary{decorations.pathreplacing,decorations.markings}

\usepackage{booktabs}

\usepackage{caption}
\usepackage{subcaption}

\usepackage{quiver}
\usepackage{epsdice}
\usepackage{listings}
\usepackage{epigraph}
\newcommand{\indep}[0]{\!\perp\!\!\!\perp}

\def\ie{{i.e.,}\ }
\def\eg{{e.g.,}\ }
\newcommand\vcdice[1]{\vcenter{\hbox{\epsdice{#1}}}}

\Title{Higher-Order Interactions and their Duals Reveal Synergy and Logical Dependence Beyond Shannon-Information}
\TitleCitation{Higher-Order Interactions and their Duals Reveal Synergy and Logical Dependence Beyond Shannon-Information}
 
\Author{Abel Jansma $^{1,2,3}$\orcidA{}}
 
\AuthorNames{Abel Jansma}
 
\AuthorCitation{Jansma, A.}
 
\address{%
$^{1}$ \quad MRC Human Genetics Unit, Institute of Genetics \& Cancer, University of Edinburgh, \mbox{Edinburgh EH8 9YL, UK;} A.A.A.Jansma@sms.ed.ac.uk \\
$^{2}$ \quad Higgs Centre for Theoretical Physics, School of Physics \& Astronomy, University of Edinburgh, \mbox{Edinburgh EH8 9YL, UK} \\
$^{3}$ \quad Biomedical AI Lab, School of Informatics, University of Edinburgh, \mbox{Edinburgh EH8 9YL, UK} }
\abstract{Information-theoretic quantities reveal dependencies among variables in the structure of joint, marginal, and conditional entropies while leaving certain fundamentally different systems indistinguishable. Furthermore, there is no consensus on the correct higher-order generalisation of mutual information (MI). In this manuscript, we show that a recently proposed model-free definition of higher-order interactions among binary variables (MFIs), like mutual information, is a Möbius inversion on a Boolean algebra, except of surprisal instead of entropy. This provides an information-theoretic interpretation to the MFIs, and by extension to Ising interactions. We study the objects dual to mutual information and the MFIs on the order-reversed lattices. We find that dual MI is related to the previously studied differential mutual information, while dual interactions are interactions with respect to a different background state. Unlike (dual) mutual information, interactions and their duals uniquely identify all six 2-input logic gates, the dy- and triadic distributions, and different causal dynamics that are identical in terms of their Shannon information content.}


\begin{document}

\section{Introduction \label{sec:intro}}
\unskip
\subsection{Higher-Order~Interactions} 
All non-trivial structures in data or probability distributions correspond to dependencies among the different features, or~variables. These dependencies can be present among pairs of variables, {\ie} pairwise, or can be \textit{higher-order}. A~dependency, or~interaction, is called higher-order if it is inherently a property of more than two variables and~if it cannot be decomposed into pairwise quantities. The~term has been used more generally to refer simply to complex interactions, as for example in~\cite{Ghazanfar2020} to refer to changes in gene co-expression over time;~in this article, however, it is used only in the stricter sense defined in Section~\ref{sec:background}.

The reason such higher-order structures are interesting is twofold. First, higher-order dependence corresponds to a fundamentally different kind of communication and interaction among the components of a system. If~a system contains higher-order interactions, then its dependency structure cannot be represented by a graph and~requires a hypergraph, where a single `hyperedge' can connect more than two nodes. It is desirable to be able to detect and~describe such systems accurately, which requires a good understanding of higher-order interactions. Second, higher-order interactions might play an important role in nature, and~have been identified in various interaction networks, including genetic~\cite{Lezon2006, Watkinson2009, Kuzmin2018, Weinreich2013}, neuronal~\cite{Panas2013, Tkacik2014, Ganmor2011, Yu2011, Gatica2021}, ecological~\cite{Sanchez2019, Grilli2017, Li2021}, drug interaction~\cite{Tekin2018}, social~\cite{Alvarez2021, Cencetti2021, Grabisch1999}, and~physical~\cite{Matsuda2000, Cerf1997} networks. Furthermore, there is evidence that higher-order interactions are responsible for the rich dynamics~\cite{Battiston2021} or bistability~\cite{Skardal2020} in biological networks; for example, synthetic lethality experiments have shown that the trigenic interactions in yeast form a larger network than the pairwise interactions~\cite{Kuzmin2018}. 

Despite this, purely pairwise descriptions of nature have been remarkably successful, which the authors of~\cite{suffPairwise, Tkacik2006} attribute to the fact that there are regimes in terms of the strength and density of coupling among the variables within which pairwise descriptions are sufficient. Alternatively, it may be attributed to the fact that higher-order interactions have been understudied and~their effects underestimated. Currently, perhaps the most promising method of quantifying higher-order interactions is information theory. The~two most commonly used quantities are mutual information and its higher-order generalisation (used in, {\eg}  
\cite{Margolin2006, Nemenman2004}) and~the total correlation (introduced in~\cite{Watanabe1960} and recently used in~\cite{Rosas2022}). However, one particular problem of interest that total correlation and mutual information do not address is that of synergy and redundancy. Given a set of variables with an $n$th-order dependency, what part of that is exclusively $n$th-order (called the synergistic part), and~what part can be found in a subset of $m<n$ variables as well (the redundant part)? Quantifying the exact extent to which  shared information is synergistic is an open problem, and is~most commonly addressed using \textit{partial information decomposition} \citep{PID}, which has been applied mainly in the context of theoretical neuroscience \citep{Wibral2017}. In~this article, a~different more statistical approach to identifying synergy is taken, which is ultimately shown to be intimately related to information theory while~offering significant advantages beyond classical entropy-based~quantities.

\subsection{Model-Free Interactions and the Inverse Ising~Problem} 
In 1957, E.T. Jaynes famously showed that statistical equilibrium mechanics can be seen as a maximum entropy solution to the inverse problem of constructing a probability distribution that best reproduces a sample distribution~\cite{Jaynes_maxEnt}. More precisely, the~equilibrium dynamics of the (inhomogeneous, or~glass-like) generalised Ising model with interactions up to the $n$th order arise naturally as the maximum entropy distribution compatible with a dataset after observing the first $n$ moments among binary variables. This means that in order to reproduce the moments in the data in a maximally non-committal way, it is necessary to introduce higher-order interactions, \ie terms that involve more than two variables, in~the description of the system. Fitting such a generalised Ising model to data is nontrivial; while the log-likelihood of the Ising model is concave in the the coupling parameters, the~cost of evaluating it is exponential in the total number of variables $N$, which is often intractable in practice~\cite{Nguyen2017}. In~\cite{AvaSjoerd}, the~authors introduced an estimator of model-free interactions (MFIs) that exactly coincides with the solution to the inverse generalised Ising problem. Moreover, the~cost of estimating all $n$th-order model-free interactions among $N$ variables from $M$ observations scales as $\mathcal{O}\left( M \cdot { N \choose n}\right) = \mathcal{O}(M N^n)$ (i.e.,~polynomially) in~the total system size $N$. However, this is true only when sufficient data is available. With~limited data, certain interactions might require inferring the conditional dependencies from the data, which in the worst case scales exponentially in $N$ again. The~definition of MFIs offered in~\cite{AvaSjoerd} seems to be a general one; in addition to offering a solution to the inverse generalised Ising problem,~MFIs are expressible in terms of average treatment effects (ATEs) or regression coefficients. Throughout this article, the~general term `MFI' is used, and~may be read simply as referring to the maximum entropy or~Ising interaction.

\subsection{Outline}
In Section~\ref{sec:MFIdef}, the~definition of the MFIs is stated along with a number of their properties. To~explicitly link the MFIs to information theory, a~redefinition of mutual information in terms of Möbius inversions is provided in Section~\ref{sec:MIasMI}, which is then linked to a similar redefinition of the MFIs in Sections~\ref{sec:MFIasMI} and \ref{sec:catMFIs}. A~definition in terms of Möbius inversions naturally leads to dual definitions of all objects, which are subsequently explored in Section~\ref{sec:dualLattices}. Then, in~Section~\ref{sec:results}, simple fundamental examples are used to demonstrate that MFIs can differentiate distributions that entropy-based quantities cannot. Finally, the~results are summarised and reflected upon in Section~\ref{sec:discussion}.

\section{Background \label{sec:background}}
\unskip
\subsection{Model-Free~Interactions \label{sec:MFIdef}}

We start by re-defining the interactions introduced in~\cite{AvaSjoerd}. We define the isolated effect (or~1-point interaction) $I^{(Y)}_i$ of a variable $X_i \in X$ on an observable $Y$ as
\begin{align}
	I^{(Y)}_{i} &= \frac{\partial Y}{\partial X_i} \Big|_{\underline{X}=0}~~~,~~~ \underline{X} = X \setminus \{X_i\}\\
	\intertext{where the effect of $X_i$ on $Y$ is isolated by conditioning on all other variables being zero. This expression is well-defined, as the restriction of a derivative is the derivative of the restriction. A~pair of variables $X_i$ and $X_j$ has a 2-point interaction $I^{(Y)}_{ij}$ when the value of $X_j$ changes the 1-point interaction of $X_i$ on $Y$:}
	I^{(Y)}_{ij} &= \frac{\partial I^{(Y)}_{i}}{\partial X_j}\Big|_{\underline{X}=0} =  \frac{\partial ^2 Y }{\partial X_j \partial X_i }\Big|_{\underline{X}=0}~~~,~~~ \underline{X} = X \setminus \{X_i, X_j\}
	\intertext{A third variable $X_k$ can modulate this 2-point interaction through what we call a 3-point interaction, $I^{(Y)}_{ijk}$:}
	I^{(Y)}_{ijk} &= \frac{\partial I^{(Y)}_{ij}}{\partial X_k}\Big|_{\underline{X}=0} =  \frac{\partial ^3 Y }{\partial X_k \partial X_j  \partial X_i}\Big|_{\underline{X}=0}~~~,~~~ \underline{X} = X \setminus \{X_i, X_j, X_k\}
\end{align}
This process of taking derivatives with respect to an increasing number of variables can be repeated to define $n$-point interactions.

\begin{defn}[$n$-point interaction with respect to outcome $Y$]\label{def:intWrtOutcome}
	Let p be a probability distribution over a set $X$ of random variables $X_i$ and let $Y$ be a function $Y:X\to \mathbb{R}$. Then, the $n$-point interaction $I_{X_1\ldots X_n}$ between variables $\{X_1, \ldots, X_n\} \subseteq X$ is provided by
\begin{align}
  		I^{(Y)}_{X_1\ldots X_n} = \frac{\partial^n Y(X)}{\partial X_1\ldots \partial X_n} \Big|_{\underline{X}=0}
	\end{align}
	where $\underline{X} = X \setminus \{X_1, \ldots X_n\}$.
\end{defn}

This definition of interaction makes explicit the fact that interactions are defined with respect to some outcome. The~authors of~\cite{AvaSjoerd} refer to the interactions from Definition \ref{def:intWrtOutcome} as \textit{additive}, which they distinguish from \textit{multiplicative} interactions. However, when the outcome is chosen to be the log of the joint distribution $p(X)$ over all variables $X$, then the additive and multiplicative interactions are equivalent and simply related through a logarithm~\cite{AvaSjoerd}. Setting the outcome to be $\log p(X)$ has other nice properties as well. First, while probabilities are restricted to the non-negative reals,~a log-transformation removes this restriction and makes the outcome and subsequent interactions take both positive and negative values, which can have different interpretations. Second, it is this outcome that makes the interactions interpretable as maximum entropy interactions, as~they exactly coincide with Ising interactions. Finally, this can be considered the most general outcome possible, as all marginal and conditional probabilities are encoded in this joint distribution. This leads to the following definition of a model-free interaction.

\begin{defn}[model-free $n$-point interaction between binary variables]\label{def:MFIs}
	A model-free $n$-point interaction (MFI) is an $n$-point interaction between binary random variables with respect to the logarithm of their joint probability
\begin{align}
  		I_{X_1\ldots X_n}\coloneqq I^{(\log p(X))}_{X_1\ldots X_n} = \frac{\partial^n \log p(X)}{\partial X_1\ldots \partial X_n} \Big|_{\underline{X}=0}
	\end{align}
	where $\underline{X} = X \setminus \{X_1, \ldots X_n\}$.
\end{defn}

If the variables $X_i \in X$ are binary, then a definition for a derivative with respect to a binary variable is needed.
\begin{defn}[derivative of a function with respect to a binary variable]\label{def:binDeriv}
	Let $f: \mathbb{B}^n \to \mathbb{R}$ be a real-valued function of a set $X$ of $n$ binary variables, labelled as $X_i$, $1 \leq i \leq n$. Then, the derivative operator with respect to $X_i$ acts on $f(X)$ as follows:
\begin{align}
		\frac{\partial}{\partial X_i} f (X) = f(X_i=1, X\setminus X_i) - f(X_i=0, X\setminus X_i)
	\end{align}
	The linearity of the derivative operator then immediately and uniquely defines the higher-order derivatives. 
\end{defn}

Using this definition, the~n-point interactions become model-free in the sense that they are ratios of probabilities that do not involve the functional form of the joint probability distribution. For~example, writing $X_{ijk}=(a, b, c)$ for $(X_i = a, X_j = b, X_k=c)$, the~first three orders can be written out as follows (recall that the notation $\frac{\partial}{\partial X_i}$ here refers to the derivative operator from Definition \ref{def:binDeriv}):
\begin{align}
	I_i &= \frac{\partial \log p(X)}{\partial X_i}\Big|_{\underline{X}=0} = \log \frac{p\Big(X_i=1 \mid \underline{X}=0\Big)}{p\Big(X_i=0 \mid \underline{X}=0\Big)} \label{eq:1-point} \\
	I_{ij} &= \frac{\partial ^2 \log p(X) }{\partial X_j \partial X_i }\Big|_{\underline{X}=0} = \log \frac{p\Big(X_{ij}=(1,1) \mid \underline{X}=0\Big)}{p\Big(X_{ij} = (0, 1) \mid \underline{X}=0\Big)} \frac{p\Big(X_{ij}=(0, 0) \mid \underline{X}=0\Big)}{p\Big(X_{ij} = (1, 0) \mid \underline{X}=0\Big)}\label{eq:2-point}\\
	\nonumber I_{ijk} &= \frac{\partial ^3 \log p(X) }{\partial X_k \partial X_j \partial X_i }\Big|_{\underline{X}=0} = \\
	 \nonumber \log &\frac{p\Big(X_{ijk}=(1, 1, 1) \mid \underline{X}=0\Big)}{p\Big(X_{ijk}=(0, 0, 0) \mid \underline{X}=0\Big)}	
		\frac{p\Big(X_{ijk}=(1, 0, 0) \mid \underline{X}=0\Big)}{p\Big(X_{ijk}=(0, 1, 1) \mid \underline{X}=0\Big)}\\
		\times &\frac{p\Big(X_{ijk}=(0, 1, 0) \mid \underline{X}=0\Big)}{p\Big(X_{ijk}=(1, 0, 1) \mid \underline{X}=0\Big)}
		\frac{p\Big(X_{ijk}=(0, 0, 1) \mid \underline{X}=0\Big)}{p\Big(X_{ijk}=(1, 1, 0) \mid \underline{X}=0\Big)} \label{eq:3-point}
\end{align}
where Bayes' rule is used to replace joint probabilities with conditional probabilities. This definition of interaction has the following~properties:
\begin{itemize}
    \item It is symmetric in terms of the variables, as $I_{S} = I_{\pi(S)}$ for any set of variables $S$ and~any permutation $\pi$. 
    \item Conditionally independent variables do not interact: $X_i \indep X_j \mid \underline{X} \implies I_{ij}=0$.
    \item If $\underline{X}=\emptyset$, the~definition coincides with that of a log-odds ratio, which has already been considered as a measure of interaction in, \eg~\cite{logOddsInteractions_1,logOddsInteractions_2}.
    \item The interactions are model-free; no knowledge of the functional form of $p(X)$ is required, and~the probabilities can be directly estimated from i.i.d. samples.
    \item The MFIs are exactly the Ising interactions in the maximum entropy model after observing moments of the data. This can be readily verified by setting $$p(s) = \mathcal{Z}^{-1} \exp (\sum_n \sum_{i_1, \ldots, i_n} J_{i_1\ldots i_n} s_{i_1}\ldots s_{i_n})$$ and using Definition \ref{def:MFIs}.
\end{itemize}

{Furthermore,}  
 in~Appendix \ref{A:proofs} the following two useful properties are introduced and~proved:
\begin{itemize}
	\item An $n$-point interaction can only be non-zero if all $n$ variables are in each other's minimal Markov blanket.
    \item If $\underline{X}$ does not include the full complement of the interacting variables, the~bias this induces in the estimate of the interaction is proportional to the pointwise mutual information of states where the omitted variables are 0. 
\end{itemize}

\subsection{Mutual Information as a Möbius~Inversion \label{sec:MIasMI}}

The definition of an $n$-point interaction as a derivative of a derivative is reminiscent of Gregory Bateson's view of information as a \textit{difference which makes a difference} \cite{Bateson1972}; however,~the relationship between information theory and model-free interactions rests on more than a linguistic coincidence. It turns out that interactions and information are generalised derivatives of similar functions on Boolean algebras. To~see this, consider the definition of pairwise mutual information and~its third-order generalisation:
\begin{align}
		MI(X, Y) &= H(X) - H(X\mid Y)\\
		&= H(X) + H(Y) - H(X, Y) \label{eq:MI_2Var}\\
		MI(X, Y, Z) &= MI(X, Y) - MI(X, Y \mid Z)\\
		\nonumber ~&=H(X)+H(Y)+H(Z) \\
		\nonumber &-H(X, Y)-H(X, Z)-H(Y, Z) + H(X, Y, Z) \label{eq:MI_3Var}
	\end{align}

Note that all MI-based quantities can be written thusly as sums of marginal entropies of subsets of the set of variables. Given a finite set of variables $S$, its powerset $\mathcal{P}(S)$ can be assigned a partial ordering as follows:
\begin{align}
	a\leq b \iff a \subseteq b ~~~~ \forall ~ a, b \in \mathcal{P}(S)
\end{align}

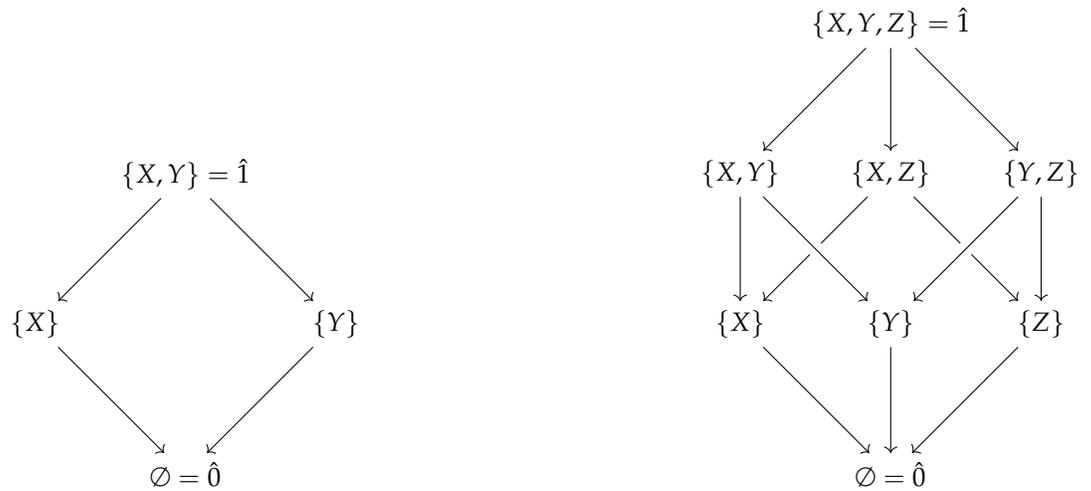
\begin{figure}[H]
\begin{tikzpicture}[baseline]
  \node[align=center] (max) at (0,2) {$\{X, Y\} = \hat{1}$};
  \node[align=center] (a) at (-2,0) {$\{X\}$};
  \node[align=center] (b) at (2,0) {$\{Y\}$};
  \node[align=center] (min) at (0,-2) {$\emptyset = \hat{0}$};
  \draw[->] (max) -- (a);
  \draw[->] (max) -- (b);
  \draw[->] (a) -- (min);
  \draw[->] (b) -- (min);
\end{tikzpicture}
\hfill
\begin{tikzpicture}[baseline]
  \node (max) at (0,4) {$\{X, Y, Z\}= \hat{1}$};
  \node (a) at (-2,2) {$\{X, Y\}$};
  \node (b) at (0,2) {$\{X, Z\}$};
  \node (c) at (2,2) {$\{Y, Z\}$};
  \node (d) at (-2,0) {$\{X\}$};
  \node (e) at (0,0) {$\{Y\}$};
  \node (f) at (2,0) {$\{Z\}$};
  \node (min) at (0,-2) {$\emptyset = \hat{0}$};
  \draw[->] (max) -- (a);
  \draw[->] (max) -- (b);
  \draw[->] (max) -- (c);
  \draw[->] (a) -- (d);
  \draw[->] (b) -- (d);
  \draw[->] (b) -- (f);
  \draw[->] (c) -- (f);
  \draw[->] (d) -- (min);
  \draw[->] (e) -- (min);
  \draw[->] (f) -- (min);
  \draw[->, preaction={draw=white, -,line width=6pt}] (a) -- (e);
  \draw[->, preaction={draw=white, -,line width=6pt}] (c) -- (e);
\end{tikzpicture}
	\caption{The lattices associated with $\mathcal{P}(\{X, Y\})$ (\textbf{left}) and $\mathcal{P}(\{X, Y, Z\})$ (\textbf{right}) ordered by inclusion. An~arrow $b\to a$ indicates $a < b$.}
	\label{fig:lattices}
\end{figure}

This poset $P = (\mathcal{P}(S), \subseteq)$ is called a Boolean algebra, and~because each pair of sets has a unique supremum (their union) and infimum (their intersection), it is a lattice. This lattice structure is visualised for two and three variables in Figure~\ref{fig:lattices}. In~general, the~lattice of an $n$-variable Boolean algebra forms an $n$-cube. Furthermore, for~any finite $n$, the~$n$-variable Boolean algebra forms a bounded lattice, which means that it has a \textit{greatest} element, denoted as $\hat{1}$, and~a \textit{least} element, denoted as $\hat{0}$.

\vspace{-6pt}

On a poset $P$, we define the Möbius function $\mu_P: P \times P \to \mathbb{R}$ as 

\begin{align}
  	\mu_P(x, y) =
    \begin{cases}
      1 & \text{if } x=y\\
      -\sum\limits_{z: x\leq z < y}\mu_P(x, z) & \text{if } x < y  \\
      0 & \text{otherwise} \label{eq:MF}
    \end{cases}       
\end{align}

{{This}
	function type makes $\mu_P$ an element of the \textit{incidence algebra} of $P$. In~fact, $\mu$ is the inverse of the zeta function $\zeta: \zeta(x, y) = 1 \text{ iff } x\leq y$, and 0 otherwise.~}
On a Boolean algebra, such as a powerset ordered by inclusion, the~Möbius function takes the simple form $\mu(x, y) = (-1)^{\mid x \mid - \mid y \mid}$ \cite{enumComb, Rota1964}.
This definition allows the mutual information among a set of variables $\tau$ to be written as follows~\cite{coInfoLattice, Galas2019a}:
\begin{align}
	MI(\tau) &= (-1)^{\mid\tau\mid -1} \sum_{\eta \leq \tau} \mu_P(\eta, \tau) H(\eta) \label{eq:MIasMTF}\\
	&=\sum_{\eta \leq \tau} (-1)^{\mid\eta\mid + 1} H(\eta) 
\end{align}
where $P$ is the Boolean algebra with $\tau = \hat{1}$ and~$H(\eta)$ is the marginal entropy of the set of variables $\eta$. Indeed, this coincides with Equation \eqref{eq:MI_2Var} for $\tau=\{X, Y\}$ and with Equation \eqref{eq:MI_3Var} for $\tau=\{X, Y, Z\}$. Equation \eqref{eq:MIasMTF} is a convolution known as a Möbius inversion.

\begin{defn}[Möbius inversion over a poset, Rota (1964)~\cite{Rota1964}]
	Let P be a poset $(S, \leq)$, let $\mu: P\times P \to \mathbb{R}$ be the Möbius function from Equation \eqref{eq:MF}, and let $g: P \to \mathbb{R}$ be a function on $P$. Then, the function\vspace{-6pt}
\begin{align}
  f(y) = \sum\limits_{x\leq y} \mu_P(x, y) g(x) \label{eq:mobiusInversion}
\end{align}
is called the Möbius inversion of $g$ on $P$. Furthermore, this equation can be inverted to yield
\begin{align}
  f(y) = \sum\limits_{x\leq y} \mu_P(x, y) g(x) \iff g(y)=\sum_{x\leq y}f(x) \label{eq:mobiusInversion_inv}
\end{align}
\end{defn}
 The Möbius inversion is a generalisation of the derivative to posets. If~\mbox{$P = (\mathbb{N}, \leq)$,} Equation \eqref{eq:mobiusInversion_inv} is just a discrete version of the fundamental theorem of calculus~\cite{enumComb}.\linebreak \mbox{Equation \eqref{eq:mobiusInversion_inv}} \textls[-15]{additionally implies that we can express joint entropy as a sum over mutual} \mbox{information}:
\begin{align}
  H(\tau) = (-1)^{\mid\tau\mid -1}\sum_{\eta \leq \tau} MI(\eta)
\end{align}
For example, in~the case of three variables,\vspace{-6pt}
\begin{adjustwidth}{-\extralength}{0cm}
\centering 
\begin{align}
	H(X, Y, Z) = MI(X,Y,Z) + MI(X, Y) + MI(X, Z) + MI(Y, Z) + H(X) + H(Y) + H(Z)
\end{align}
\end{adjustwidth}

Instead of starting with entropy, we could start with a quantity known as surprisal, or~self-information, defined as the negative log probability of a certain state or realisation:
\begin{align}
  S(X=x) = &-\log p(X=x)
\end{align}
Surprisal plays an important role in information theory;~indeed, the~expected surprisal across all possible realisations $X=x$ is the entropy of the variable $X$:
\begin{align}
  \mathbb{E}_X[S(X=x)] = H(X)
\end{align}
As we are often interested in the marginal surprisal of a realisation $X=x$ summed over $Y$, we can write this explicitly as
\begin{align}
 \log p(x;Y) \coloneqq \sum_y \log p(x, y)
\end{align}
With this, consider the Möbius inversion of the marginal surprisal over the lattice $P$:
\begin{align}
  \text{pmi}(T=\tau) \coloneqq  (-1)^{\mid\tau\mid}\sum_{\eta\leq \tau} \mu_P(\eta, \tau) \log p(\eta;\tau\setminus\eta)
\end{align}
This is a generalised version of the pointwise mutual information, which is usually defined on just two variables:
\begin{align}
  \text{pmi}(X=x, Y=y) &= \log(x, y;\emptyset) - \log(x;Y) - \log(y;X) + \log(\emptyset;X, Y)\\
   &= \log \frac{p(x, y)}{p(x)p(y)}
\end{align}

\noindent \textbf{{Summary}  
}
\begin{itemize}
	\item \textit{{Mutual}  
 information is the Möbius inversion of marginal entropy}.
	\item \textit{Pointwise mutual information is the Möbius inversion of marginal surprisal}.
\end{itemize}

\section{Interactions and Their~Duals}
\unskip
\subsection{MFIs as Möbius~Inversions \label{sec:MFIasMI}}
With mutual information defined in terms of Möbius inversions, the~same can be done for the model-free interactions. Again, we start  with~(negative) surprisal. However, on~Boolean variables a~state is just a partition of the variables into two sets: one in which the variables are set to 1, and~another in which they are set to 0. That means that the surprisal of observing a particular state is completely specified by which variables $X\subseteq Z$ are set to 1 while keeping all other variables $Z\setminus X$ at 0, which can be written as
\begin{align}
   S_{X;Z} \coloneqq &\log p(X=1, Z\setminus X =0)
\end{align}

\begin{defn}[interactions as Möbius inversions]\label{def:MFIsAsMIs}
	Let $p$ be a probability distribution over a set $T$ of random variables and let $P = (\mathcal{P}(\tau), \subseteq)$, the~powerset of a set $\tau \subseteq T$ ordered by inclusion. Then, the interaction $I(\tau;T)$ among variables $\tau$ is provided by
\begin{align}
  I(\tau;T) \coloneqq& \sum_{\eta\leq\tau} \mu_P(\eta, \tau) S_{\eta;T}\\
  =& \sum_{\eta\leq\tau} (-1)^{\mid \eta \mid - \mid \tau \mid} \log p(\eta=1, T\setminus\eta=0)
\end{align}
\end{defn}
For example, when $\tau$ contains a single variable $X\subseteq T$, then
\begin{align}
  I(\{X\};T) &= \mu_P(\{X\}, \{X\}) S_{\{X\};T} + \mu_P(\emptyset, \{X\}) S_{\emptyset;T}\\
		&= \log \frac{p(X=1, T\setminus X =0)}{p(X=0,  T\setminus X =0)}
\end{align}
which coincides with the 1-point interaction in Equation \eqref{eq:1-point}. Similarly, when $\tau$ contains two variables $\tau = \{X, Y\}\subseteq T$, then
\begin{align}
  I(\{X, Y\} ;T) = \mu_P(\{X, Y\}, \{X, Y\}) S_{\{X, Y\};T} + \mu_P(\{X\}, \{X, Y\}) S_{\{X\};T}\\
  \nonumber + \mu_P(\{Y\}, \{X, Y\}) S_{\{Y\};T} + \mu_P(\emptyset, \{X, Y\}) S_{\emptyset;T}\\
= \log \frac{p(X=1, Y=1, T\setminus \{X, Y\} =0) p(X=0, Y=0, T\setminus \{X, Y\} =0)}{p(X=1, Y=0, T\setminus \{X, Y\} =0) p(X=0, Y=1, T\setminus \{X, Y\} =0)}
\end{align}
which coincides with the 2-point interaction in Equation \eqref{eq:2-point}. In~fact, this pattern holds in general.

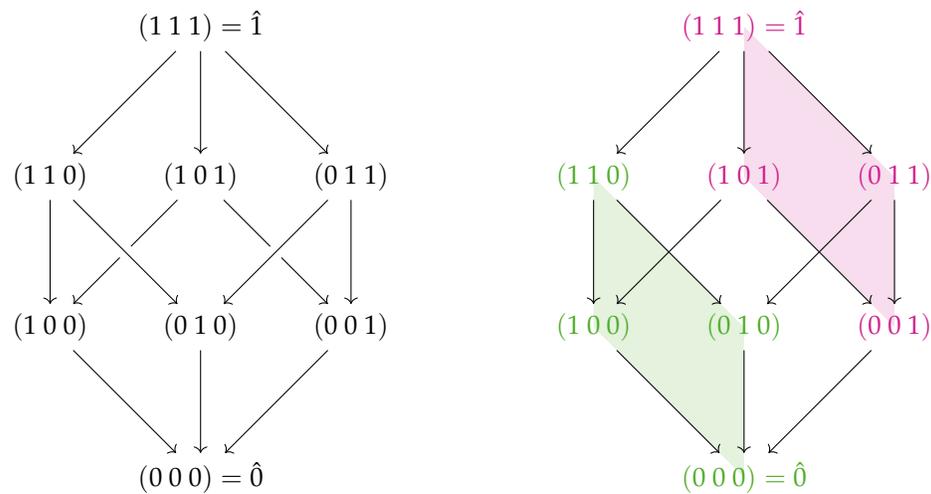
\begin{figure}[H]
\begin{tikzpicture}
  \node (max) at (0,4) {$(1~1~1) = \hat{1}$};
  \node (a) at (-2,2) {$(1~1~0)$};
  \node (b) at (0,2) {$(1~0~1)$};
  \node (c) at (2,2) {$(0~1~1)$};
  \node (d) at (-2,0) {$(1~0~0)$};
  \node (e) at (0,0) {$(0~1~0)$};
  \node (f) at (2,0) {$(0~0~1)$};
  \node (min) at (0,-2) {$(0~0~0) = \hat{0}$};
  \draw[->] (max) -- (a);
  \draw[->] (max) -- (b);
  \draw[->] (max) -- (c);
  \draw[->] (a) -- (d);
  \draw[->] (b) -- (d);
  \draw[->] (b) -- (f);
  \draw[->] (c) -- (f);
  \draw[->] (d) -- (min);
  \draw[->] (e) -- (min);
  \draw[->] (f) -- (min);
  \draw[->, preaction={draw=white, -,line width=6pt}] (a) -- (e);
  \draw[->, preaction={draw=white, -,line width=6pt}] (c) -- (e);
\end{tikzpicture}
\hfil
\begin{tikzpicture}
  \node [cbRed] (max) at (0,4) {$(1~1~1) = \hat{1}$};
  \node [cbGreen] (a) at (-2,2) {$(1~1~0)$};
  \node [cbRed] (b) at (0,2) {$(1~0~1)$};
  \node [cbRed] (c) at (2,2) {$(0~1~1)$};
  \node [cbGreen](d) at (-2,0) {$(1~0~0)$};
  \node [cbGreen](e) at (0,0) {$(0~1~0)$};
  \node [cbRed] (f) at (2,0) {$(0~0~1)$};
  \node [cbGreen](min) at (0,-2) {$(0~0~0) = \hat{0}$};
  \draw[->] (max) -- (a);
  \draw[->] (max) -- (b);
  \draw[->] (max) -- (c);
  \draw[->] (a) -- (d);
  \draw[->] (a) -- (e);
  \draw[->] (b) -- (d);
  \draw[->] (b) -- (f);
  \draw[->] (c) -- (f);
  \draw[->] (c) -- (e);
  \draw[->] (d) -- (min);
  \draw[->] (e) -- (min);
  \draw[->] (f) -- (min);
  \begin{pgfonlayer}{bg}
    \fill[cbRed, opacity=.15] (max.center) to [smooth] (c.center)
                   to [smooth] (f.center)
                   to [smooth] (b.center);
    \fill[cbGreen, opacity=.15] (a.center)
                   to (d.center)
                   to (min.center)
                   to (e.center);
  \end{pgfonlayer}
\end{tikzpicture}
\hfil
	\caption{(\textbf{{Left} 
}) The lattice associated with $\mathcal{P}(\{X, Y, Z\})$ ordered by inclusion as~binary strings. Equivalently, the~lattice of binary strings where for any two strings $a$ and $b$, $a\leq b \iff a\land b = a$. (\textbf{Right}): The two shaded regions correspond to the decomposition of the 3-point interaction into two 2-point~interactions.}
	\label{fig:lattices_strings}
\end{figure}

\begin{Theorem}[equivalence of interactions]
	The interaction $I(\tau, T)$ from Definition \ref{def:MFIsAsMIs} is the same as the model-free interaction $I_\tau$ from Definition \ref{def:MFIs}, that is, for~any set of variables $\tau \subseteq T$ it is the case that
\begin{align}
  I(\tau, T) = I_\tau
\end{align}
\end{Theorem}
\begin{proof}
	We have to show that
\begin{align}
  \sum_{\eta\leq\tau} (-1)^{\mid \eta \mid - \mid \tau \mid} \log p(\eta=1, T\setminus\eta=0) = \frac{\partial^n \log p(T)}{\partial \tau_1\ldots \partial \tau_n} \Big|_{\underline{T}=0} \label{eq:toShow}
\end{align}

Both sides of this equation are sums of $\pm\log p(s)$, where $s$ is some binary string; thus, we have to show that the same strings appear with the same~sign.

First, note that the Boolean algebra of sets ordered by inclusion (as in Figure~\ref{fig:lattices}) is equivalent to the poset of binary strings where for any two strings $a$ and $b$, $a\leq b \iff a\land b = a$. The~equivalence follows immediately upon setting each element $a \in \mathcal{P}(S)$ to the string where $a=1$ and $S\setminus a = 0$. This map is one-to-one and monotonic with respect to the partial order, as $A \subseteq B \iff A \cap B = A$. This means that Definition \ref{def:MFIsAsMIs} can be rewritten as a Möbius inversion on the lattice of Boolean strings $S = (\mathbb{B}^{|\tau|}, \leq)$ (shown for the three-variable case on the left side of Figure~\ref{fig:lattices_strings}):
\begin{align}
  I(\tau;T) =& \sum_{s \leq \hat{1}_S} \mu_S(s, \hat{1}_S) \log p(\tau=s, T\setminus \tau=0) \label{eq:MFIs_strings}\\
  \intertext{Note that for any pair ($\alpha, \tau)$ where $\alpha\subseteq \tau$ with~respective string representations $(s, t) \in \mathbb{B}^{|\tau|} \times \mathbb{B}^{|\tau|}$, we have the following:}
  |\tau| - |\alpha| =& \sum_i (t \land \lnot s)_i\\
  \intertext{Thus, we can write}
  I(\tau;T) =& \sum_{s \leq \hat{1}_S} (-1)^{\sum \lnot s} \log p(\tau=s, T\setminus \tau=0) \label{eq:interactionsOnBooleanStrings}
\end{align}
To see that this exactly coincides with Definition \ref{def:MFIs}, we can define a map
\begin{align}
	e^{(n)}_{i, s}: \mathcal{F}_{\mathbb{B}^n} &\to \mathcal{F}_{\mathbb{B}^{n-1}}
	\intertext{where $\mathcal{F}_{\mathbb{B}^n}$ is the set of functions from $n$ Boolean variables to $\mathbb{R}$. This map is defined as}
	e^{(n)}_{i, s}: f(X_1, \ldots X_i, \ldots X_n) &\mapsto f(X_1, \ldots X_i=s, \ldots X_n)
	\intertext{With this map, the~Boolean derivative of a function $f(X_1, \ldots,  X_n)$ (see Definition \ref{def:binDeriv}) can be written as}
	\frac{\partial}{\partial X_i} f(X) &= (e^{(n)}_{i, 1} - e^{(n)}_{i, 0}) f(X) \\
	= f(X_1, \ldots, X_i=1, \ldots, X_n) &- f(X_1, \ldots, X_i=0, \ldots, X_n)
	\intertext{In this way, the derivative with respect to a set $S$ of $m$ variables becomes function composition:}
	\left(\prod_{i=0}^m\frac{\partial}{\partial X_{S_i}}\right) f(X) &= \left(\bigcirc_{i=0}^m (e^{(n-i)}_{S_i, 1} - e^{(n-i)}_{S_i, 0})\right) f(X)
\end{align}

From this, it is clear that a term $f(s)$ appears with a minus sign iff $e^{(n)}_{i, 0}$ has been applied an odd number of times. Therefore, terms for which $s$ contains an odd number of $0$s receive a minus sign. This can be summarised as
\begin{align}
  \left(\prod_{i=0}^m\frac{\partial}{\partial X_{S_i}}\right) f(X) = \sum_{s \in \mathbb{B}^n} (-1)^{\sum \neg s} f(X_S=s, X\setminus X_S)
\end{align}
Therefore, we can write
\begin{align}
  I_\tau = \sum_{s \in \mathbb{B}^n} (-1)^{\sum \neg s} \log(\tau=s, T\setminus \tau=0)\label{eq:IasSumOfStrings}
\end{align}
The sums $\sum_{s\leq \hat{1}_S}$ and $\sum_{s \in \mathbb{B}^n}$ contain exactly the same terms, meaning that \mbox{Equations}~(\ref{eq:interactionsOnBooleanStrings}) and (\ref{eq:IasSumOfStrings}) are equal. This~completes the proof.
\end{proof}

 Note that the structure of the lattice reveals structure in the interactions, as previously noted in~\cite{AvaSjoerd}. On~the right-hand side of Figure~\ref{fig:lattices_strings}, two faces of the three-variable lattice are shaded. The~green region corresponds to the 2-point interaction between the first two variables. The~red region contains a similar interaction between the first two variables, except this time in the context of the third variable fixed to 1 instead of 0. This illustrates the interpretation of a 3-point interaction as the difference in two 2-point interactions ($I_{XYZ} = I_{XY\mid Z=1} - I_{XY \mid Z=0}$; note that $I_{XY \mid Z=0}$ is usually written as just $I_{XY}$). The~symmetry of the cube reveals the three different (though equivalent) choices as to which variable to set to 1. Treating the Boolean algebra as a die, where the sides facing up are $\epsdice{1}$, $\epsdice{2}$, and~$\epsdice{3}$, we have
\begin{align}
  I_{XYZ} = \vcdice{1} - \vcdice{6} = \vcdice{2} - \vcdice{5} = \vcdice{3} - \vcdice{4}
\end{align}

As before, we can invert Definition \ref{def:MFIsAsMIs} and express the surprise of observing a state with all ones in terms of interactions, as follows:
\begin{align}
  \log p(\tau = 1, T\setminus \tau =0) = \sum_{\eta \leq \tau} I(\eta, T)
  \intertext{For example, in~the case where $T=\{X, Y, Z\}$ and $\tau = \{X, Y\}$}
  S(1, 1, 0) = -\log p(1, 1, 0) = -I_{XY} - I_X - I_Y - I_\emptyset
\end{align}
which illustrates that when $X$ and $Y$ tend to be off ($I_X<0$ and $I_Y<0$) and~$X$ and $Y$ tend to be different ($I_{XY}<0$), observing the state (1, 1, 0) is very~surprising.

\subsection{Categorical~Interactions \label{sec:catMFIs}} 
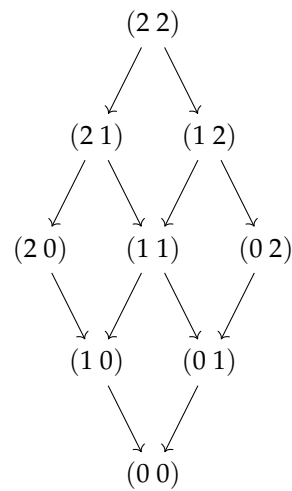
\begin{figure}[H]
\begin{tikzpicture}[scale=1.5]
  \node (max) at (0,3) {$(2~2)$};
  \node (a) at (-0.5,2) {$(2~1)$};
  \node (b) at (0.5,2) {$(1~2)$};
  \node (c) at (-1,1) {$(2~0)$};
  \node (d) at (0,1) {$(1~1)$};
  \node (e) at (1,1) {$(0~2)$};
  \node (f) at (-0.5,0) {$(1~0)$};
  \node (g) at (0.5,0) {$(0~1)$};
  \node (min) at (0,-1) {$(0~0)$};
  \draw[->] (max) -- (a);
  \draw[->] (max) -- (b);
  \draw[->] (a) -- (c);
  \draw[->] (a) -- (d);
  \draw[->] (b) -- (d);
  \draw[->] (b) -- (e);
  \draw[->] (c) -- (f);
  \draw[->] (d) -- (f);
  \draw[->] (d) -- (g);
  \draw[->] (e) -- (g);
  \draw[->] (f) -- (min);
  \draw[->] (g) -- (min);
\end{tikzpicture}
	\caption{The lattice of two variables that can take three values, ordered by $a \leq b \iff  \forall i$$: a_i \leq b_i $. }
	\label{fig:lattice_cats}
\end{figure}

Taking seriously the definition of interactions as the Möbius inversion of surprisal, one might ask what happens when surprisal is inverted over a different lattice instead of using a Boolean algebra. One example is shown in Figure~\ref{fig:lattice_cats};~it corresponds to variables that can take three values---0, 1, or~2---where states are ordered by $a \leq b \iff  \forall i: a_i \leq b_i $. To~calculate interactions on this lattice, we need to know the value of Möbius functions of type $\mu(s, 22)$. It can be readily verified that most Möbius functions of this type are zero, with the exceptions~of $\mu(22,22)=\mu(11,22)=1$ and~$\mu(21,22)=\mu(12,22)=-1$, which provide the exact terms in the interactions between two categorical variables changing from $1\to2$ (as~defined in \citep{AvaSjoerd}). Calculating interactions on different sublattices with $\hat{1}=(21), (12)$ or~$(11)$ provides us with the other categorical interactions. The~transitivity property of the interactions, i.e.,~$I(X:0\to 2, Y:0 \to 1) = I(X:0\to 1, Y:0 \to 1) + I(X:1\to 2, Y:0 \to 1)$, follows immediately from the structure of the lattice in Figure~\ref{fig:lattice_cats} and~the alternating signs of the Möbius functions on a Boolean~algebra.

\unskip

\unskip

\subsection{Information and Interactions on Dual~Lattices \label{sec:dualLattices}}
Lattices have the property that a set with the reverse order remains a lattice; that is, if~$\mathcal{L} = (S, \leq)$ is a lattice, then $\mathcal{L}^\text{op} = (S, \preceq)$ (where $\forall a, b \in S: a\preceq b \iff a\geq b$) is a lattice. This raises the question of what corresponds to mutual information and interaction on such dual lattices. {Recognising
 that a poset $\mathcal{L} = (S, \leq_\mathcal{L})$ is a category $\mathcal{C}$ with objects $S$ and a morphism $f: A \to B$ iff $B\leq_\mathcal{L}A$, these become definitions in the \textit{opposite} category $\mathcal{C}^\text{op}$, meaning that~they define dual objects.}

Let us start with mutual information. We can calculate the dual mutual information, denoted $MI^*$, by~first noting that the dual to a Boolean algebra is another Boolean algebra, meaning that we have $\mu(x, y) = (-1)^{|x| - |y|}$. Simply replacing $P$ with $P^\text{op}$ in Equation \eqref{eq:MIasMTF} yields
\begin{align}
	MI^*(\tau) &= \sum_{\eta \preceq \tau} (-1)^{|\eta| + 1} H(\eta)
\end{align}
The dual mutual information of $\tau=\hat{1}_{P^\text{op}}$ is simply $MI^*(\emptyset) = MI(\hat{1}_{P})$, that is, the~mutual information among all variables. However, the~dual mutual information of a singleton set $X$ is
\begin{align}
	MI^*(X) &= MI(\hat{1}_{P}) - MI(\hat{1}_{P}\setminus X)\\
	&=  \Delta(X; \hat{1}_{P} \setminus X)
\end{align}
where $\Delta$ is known as the differential mutual information and describes the change in mutual information when leaving out $X$ \cite{Galas2014}, \ie when marginalising over the variable $X$. Note that a similar construction was already anticipated in~\cite{Galas2019b} and that the differential mutual information has previously been used to describe information structures in genetics~\citep{Galas2019a}. On~the Boolean algebra of three variables $\{X, Y, Z\}$, the~dual mutual information of $X$ can be written out as follows:
\begin{align}
 \nonumber  MI^*(X) = \mu(\{X\}, \{X\}) H(X) + \mu(\{X, Y\}, \{X\}) H(X, Y) + \\
  \mu(\{X, Z\}, \{X\}) H(X, Z) + \mu(\{X, Y, Z\}, \{X\}) H(X, Y, Z)\\
  = H(X) - H(X, Y) - H(X, Z) + H(X, Y, Z)
\end{align}
Because $\Delta$ is the dual of mutual information, it should arguably be called the mutual co-information; however,~the term co-information is unfortunately already in use to refer to normal higher-order mutual~information.

To find the dual to the interactions, we start from Equation \eqref{eq:MFIs_strings} and construct $S^\text{op} = (\mathbb{B}^{|\tau|}, \preceq)$, the dual to the lattice of binary strings $S = (\mathbb{B}^{|\tau|}, \leq)$. A~dual interaction of variables $\tau \subseteq T$ is denoted as $I^*(\tau;T)$, and~is defined as follows:
\begin{align}
  I^*(\tau;T) &\coloneqq \sum_{s \preceq \hat{1}_{S^\text{op}}} \mu_{S^\text{op}}(s, \hat{1}_{S^\text{op}}) \log p(\tau=s, T\setminus \tau=0)\\
  \intertext{Again, when $\tau=\hat{1}_{S\text{op}} = \hat{0}_S=\emptyset$, this is simply ($-1)^{\mid \tau\mid}I(\hat{1}_S)$, while~the dual interaction of a singleton set $X$ is}
	I^*(X; T) &= (-1)^{|\hat{1}_S| - 1} \Big(I(\hat{1}_S ; T) +I(\hat{1}_S \setminus X; T)\Big)
\end{align}
For example, on~the three variable lattice in Figure~\ref{fig:lattices_strings}, the~dual interaction of $X$ is
\begin{align}
	I^*(X; T) &= I(X, Y, Z; T) + I(Y, Z; T)\\
	\intertext{Writing $p_{ijk}$ for $p(X=i, Y=j, Z=k \mid T\setminus \{X, Y, Z\}=0)$, it can be seen that this is equal to}
	I^*(X; T) &= \log \frac{p_{111} p_{100}}{p_{101} p_{110}}
\end{align}
which is similar to the 2-point interaction $I_{YZ}$ defined in Equation \eqref{eq:2-point}, now~conditioned on $X=1$ instead of 0. Note the difference between dual mutual information and dual interactions here; the dual mutual information of $X$ describes the effect on the mutual information from \textit{marginalising} over $X$, whereas the dual interaction of $X$ describes the effect on an interaction when \textit{fixing} $X=1$. This reflects a fundamental difference between mutual information and the interactions, in that the former is an \textit{averaged} quantity and the~latter a \textit{pointwise} quantity. 

Dual interactions should probably be called co-interactions; however,~to avoid confusion with the term co-information, we instead refer to them simply as dual interactions. Dual interactions are interactions that are conditioned on certain variables being 1 instead of 0. This makes them no longer equal to the Ising interactions between Boolean variables; however,~there are situations in which an interaction is more interesting in the context of $Z=1$ instead of $Z=0$, for~example, if $Z$ is always 1 in the data under~consideration. 

\noindent \textbf{{Summary} 
}
\begin{itemize}
	\item \textit{{Mutual}  
 information is the Möbius inversion of marginal entropy on the lattice of subsets ordered by inclusion.}
	\item \textit{Differential (or conditional) mutual information is the Möbius inversion of marginal entropy on the dual lattice.}
	\item \textit{Model-free interactions are the Möbius inversion of surprisal on the lattice of subsets ordered by inclusion.}
	\item \textit{Model-free dual interactions are the Möbius inversion of surprisal on the dual lattice.}
	\item \textit{Dual interactions of a variable $X$ are interactions between the other variables where $X$ is set to 1 instead of 0.}
\end{itemize}

To summarise these relationships diagrammatically, note that surprisals form a vector space as follows. Let $\mathcal{P}(T)$ be the powerset of a set of variables $T$ and~let $|\mathcal{P}(T)|=n$. This forms a lattice $P = (\mathcal{P}(T), \subseteq)$ ordered by inclusion, meaning that $\mathcal{P}(T)$ can be assigned a topological ordering indexed by $i$ as $\mathcal{P}(T) = \cup_{i=0}^n t_i$. Let $\mathcal{S}$ be the set of linear combinations of surprisals of subsets of T:
\begin{align}
  \mathcal{S} = \left\{\sum_{i=0}^n a_i \log p(t_i) \mid a_i \in \mathbb{R}\right\}
\end{align}
This set is assigned a vector space structure over $\mathbb{R}$ by the usual scalar multiplication and addition. Note that the set
\begin{align}
  \mathcal{B} = \left\{\log p(t) \mid t \in \mathcal{P}(T) \right\} 
\end{align}
forms a basis for this vector space, because $\sum_i \alpha_i \log p(t_i)=0$ has no non-trivial solutions and a~$\text{span}(\mathcal{B})=\mathcal{S}$. {Only
 when two variables $a$ and $b$ are independent do we have linear dependencies in $\mathcal{B}$, as it is~then the case that $\log p(a, b) = \log p(a) + \log p(b)$.} To~define a map from $\mathcal{S}\to \mathbb{R}$, we only need to specify its action on $\mathcal{B}$ and~extend the definition linearly. This means that we can fully define the map $ \mathit{eval}_T:\mathcal{S} \to \mathbb{R}$ by specifying
\begin{align}
   \mathit{eval}_T: \log p(R=r) &\mapsto \log p(R=1, T\setminus R=0)
\end{align}
Similarly, we can define the expectation map $\mathbb{E}: \mathcal{S} \to \mathbb{R}$ as
\begin{align}
  \mathbb{E}: \log p(R=r) &\mapsto \sum_{r} p(R=r)\log p(R=r)
\end{align}
which outputs the expected surprise over all realisations $R=r$. 
Finally, note that the Möbius inversion over a poset $P$ is an endomorphism of the set $\mathcal{F}_P$ of functions over $P$, defined as
\begin{align}
  M_P&: \mathcal{F}_P \to \mathcal{F}_P\\
  M_P&: f(y) \mapsto \sum_{x\leq y}\mu(x, y) f(x)
\end{align}
Together, these three maps ensure that the following diagram commutes:

\[\begin{tikzcd}
	{MI^*(R) = \Delta(R; \hat{1}_P)} && {H(R)} && {MI(R)} \\
	&& {} \\
	{\text{pmi}^*(R=r)} && {S(R=r;T)} && {\text{pmi}(R=r)} \\
	\\
	{I^*(R;T)} && {S_{R;T}} && {I(R;T)}
	\arrow[maps to, "{M_P}", from=1-3, to=1-5]
	\arrow[maps to, "{\mathbb{E}}"', from=3-3, to=1-3]
	\arrow[maps to, "{\mathbb{E}}"', from=3-5, to=1-5]
	\arrow[maps to, "{M_P}", from=3-3, to=3-5]
	\arrow[maps to, "{M_P}", from=5-3, to=5-5]
	\arrow[maps to, "{\textit{eval}_{T}}", from=3-3, to=5-3]
	\arrow[maps to, "{\textit{eval}_{T}}", from=3-5, to=5-5]
	\arrow[maps to, "{M_{P^{op}}}"', from=3-3, to=3-1]
	\arrow[maps to, "{M_{P^{op}}}"', from=1-3, to=1-1]
	\arrow[maps to, "{M_{P^{op}}}"', from=5-3, to=5-1]
	\arrow[maps to, "{\mathbb{E}}"', from=3-1, to=1-1]
	\arrow[maps to, "{\textit{eval}_{T}}", from=3-1, to=5-1]
\end{tikzcd}\]

For the case where $T = \{X, Y, Z\}$ and $R = \{X, Y\}$, this explicitly amounts to\vspace{-12pt}

\begin{adjustwidth}{-\extralength}{0cm}
\centering 
\[\begin{tikzcd}
\hspace{-2cm}
	{\substack{\sum_{(x, y, z) \in \mathcal{X}\times \mathcal{Y}\times\mathcal{Z}} p(x, y, z) \log p(x, y, z)\\- \sum_{(x, y) \in \mathcal{X}\times \mathcal{Y}} p(x, y)}} && {\sum_{(x, y) \in X\times Y} p(x, y) \log p(x, y)} && {\substack{\sum_{(x, y) \in \mathcal{X}\times \mathcal{Y}} p(x, y) \log p(x, y) \\- \sum_{x \in \mathcal{X}} p(x) \log p(x)\\- \sum_{y \in \mathcal{Y}} p(y) \log p(y)}} \\
	&& {} \\
	{\log\frac{p(x, y, z)}{p(x, y)}} && {\log p(x, y)} && {\log\frac{p(x, y) p(\emptyset)}{p(x)p(y)}} \\
	\\
	{\log\frac{p(1, 1, 1)}{p(1, 1, 0)}} && {\log p(1, 1, 0)} && {\log\frac{p(1, 1,0) p(0, 0,0)}{p(1, 0,0)p(0, 1,0)}}
	\arrow[maps to, "{M_P}", from=1-3, to=1-5]
	\arrow[maps to, "{\mathbb{E}}"', from=3-3, to=1-3]
	\arrow[maps to, "{\mathbb{E}}"', from=3-5, to=1-5]
	\arrow[maps to, "{M_P}", from=3-3, to=3-5]
	\arrow[maps to, "{M_P}", from=5-3, to=5-5]
	\arrow[maps to, "{\textit{eval}_{T}}", from=3-3, to=5-3]
	\arrow[maps to, "{\textit{eval}_{T}}", from=3-5, to=5-5]
	\arrow[maps to, "{M_{P^{op}}}"', from=1-3, to=1-1]
	\arrow[maps to, "{M_{P^{op}}}"', from=3-3, to=3-1]
	\arrow[maps to, "{M_{P^{op}}}"', from=5-3, to=5-1]
	\arrow[maps to, "{\mathbb{E}}"', from=3-1, to=1-1]
	\arrow[maps to, "{\textit{eval}_{T}}", from=3-1, to=5-1]
\end{tikzcd}\]
\end{adjustwidth}

\section{Results and~Examples \label{sec:results}}

While mutual information and model-free interactions are related, there are several important differences in terms of how they capture dependencies. Note, for~example, that higher-order information quantities are not independent of the lower-order quantities. The~mutual information of three variables is bounded by the pairwise quantities as follows:
\vspace{-12pt}
\begin{adjustwidth}{-\extralength}{0cm}
\centering 
{{\begin{align}
\hspace{-1.4cm}
-\min \{MI(X , Y \mid Z), MI(Y , Z \mid X), MI(X , Z \mid Y)\} \leq MI(X , Y , Z) \leq \min \{MI(X , Y), MI(Y , Z), MI(X , Z)\}
\end{align}}}
\end{adjustwidth}
This means that there are no systems with zero pairwise mutual information and positive higher-order information. This is not true for the interactions. For~example, a~distribution with 3-point interactions and no pairwise interactions can trivially be constructed as ${p(X) = \mathcal{Z}^{-1} \exp \left( \sum_{ijk} J_{ijk} X_i X_j X_k \right)}$. While this distribution has 3-point interactions with strength $J_{ijk}$ for triplets $\{X_i, X_j, X_k\}$,~all pairwise interactions among $\{X_i, X_j\}$ vanish when conditioning on $X_k=0$. In~fact, any positive discrete distribution can be written as a Boltzmann distribution with an energy function that is unique up to a constant, and as such is uniquely defined by its interactions; in~other words, each interaction, at~any order, can be freely varied to define a unique and valid probability distribution, namely, the Boltzmann distribution of the corresponding generalised Ising model. Note that this is closely related to the fact that a class of neural networks known as restricted Boltzmann machines are universal approximators \citep{Freund1994, Roux2008, Montufar2011} and~exactly (though not uniquely) encode the Boltzmann distribution of a generalised Ising model in one of their layers \citep{rbmPaper, Nguyen2017}. Therefore, each distribution is uniquely determined by its set of interactions, and~should be distinguishable by them. This is famously not true for entropy-based information quantities, as~illustrated below through several~examples.

\FloatBarrier
\subsection{Interactions and Their Duals Quantify and Distinguish Synergy in Logic~Gates}

Under the assumption of a causal collider structure $A \rightarrow C \leftarrow B$, nonzero 3-point interactions $I_{ABC}$ can be interpreted as logic gates. A~positive 3-point interaction means that the numerator in Equation \eqref{eq:3-point} is larger than the denominator. Under~the sufficient (though not necessary) assumption that each term in the numerator is larger than each term in the denominator, we obtain the following truth table as $I_{ABC} \to +\infty$:

\begin{center}
	$\begin{array}{c c |c}
	A & B & C\\ 
	\hline 
	\hline
	0 & 0 & 1\\
	0 & 1 & 0\\
	1 & 0 & 0\\
	1 & 1 & 1\\
	\end{array}$
\end{center}
which describes an XNOR gate. Let $p_\mathcal{G}$ be the probability of each of the four states in the truth table for a gate $\mathcal{G}$, and~let $\epsilon_\mathcal{G}$ be the probability of all other states. Then, the 3-point interaction of an XNOR gate can be written as\vspace{-6pt}
\begin{align}
	I^\text{XNOR}_{ABC} &= \log \frac{p_\text{XNOR}^4}{\epsilon_\text{XNOR}^4}
	\intertext{Similarly, the~truth tables of AND and OR gates imply that}
	I^\text{AND}_{ABC} &= \log \frac{ \epsilon_\text{AND} ~ p_\text{AND}^3}{\epsilon_\text{AND}^3 p_\text{AND}}\\
	I^\text{OR}_{ABC} &= \log \frac{\epsilon_\text{OR}^3 p_\text{OR}}{\epsilon_\text{OR} ~ p_\text{OR}^3}
\end{align}
If we consider equally noisy gates such that $p_\mathcal{G}=p$ and $\epsilon_\mathcal{G}=\epsilon$, the~gates can be directly compared. Note that when a gate has a 3-point interaction $I$, its logical negation will have a 3-point interaction $-I$. This determines the 3-point interactions of all six non-trivial logic gates on two inputs, as summarised in Table~\ref{tab:logicGateInteractions}. The~two gates with the strongest absolute interactions, XNOR and XOR, are the only two gates that are purely synergistic, i.e.,~knowing only one of the two inputs provides no information about the output. This relationship to synergy holds for three-input gates as well. The~three-input gate with the strongest 4-point interaction has the following truth table: 
\begin{center}
	$\begin{array}{c c c |c}
	A & B & C & D\\ 
	\hline 
	\hline
	0 & 0 & 0 & 0\\
	0 & 0 & 1 & 1\\
	0 & 1 & 0 & 1\\
	1 & 0 & 0 & 1\\
	0 & 1 & 1 & 0\\
	1 & 0 & 1 & 0\\
	1 & 1 & 0 & 0\\
	1 & 1 & 1 & 1\\
	\end{array}$
\end{center}
This is a three-input XOR gate, \ie $D=(A+B+C)\mod 2$, and~is again maximally synergistic, as observing only two of the three inputs provides zero bits of information on the output. Setting this maximum 4-point interaction to $I$, the~three-input OR and AND gates receive a 4-point interaction $I/4$; thus, the hierarchies of interaction and synergy continue to~match.

\begin{table}[H]
\caption{The 3-point interactions for all two-input logic gates at equal noise level are related through $I = 4 \log \frac{p}{\epsilon}$ and~degenerate in AND$\sim$NOR and OR$\sim$NAND.  \label{tab:logicGateInteractions}}

\setlength{\tabcolsep}{36mm}  

\resizebox{\linewidth}{!}{
	\begin{tabular}{lr}
	\toprule
	$\boldsymbol{\mathcal{G}}$ &      $\boldsymbol{I}^{\boldsymbol{\mathcal{G}}}_{\textbf{\emph{ABC}}}$    \\
	\midrule
	XNOR & $I$\\
	XOR & $-I$\\
	AND & $\frac{1}{2}I$\\
	OR & $-\frac{1}{2}I$\\
	NAND & $-\frac{1}{2}I$\\
	NOR & $\frac{1}{2}I$\\
	\bottomrule
	\end{tabular}
}
\end{table}

The 3-point interactions are able to separate most two-input logic gates by sign or value, leaving only AND$\sim$NOR and OR$\sim$NAND. Mutual information has less resolving power. Assuming a uniform distribution over all four allowed states from a gate's truth table, a~brief calculation yields
\begin{align}
	&MI^\text{OR}(A, B, C) = MI^\text{AND}(A, B, C) =MI^\text{NOR}(A, B, C) = MI^\text{NAND}(A, B, C)\\
	\nonumber&= - \log \left(\frac{3^{3/4}}{4}\right) -1\approx -0.189 \nonumber\\
	&MI^\text{XOR}(A, B, C) = MI^\text{XNOR}(A, B, C) = -1
\end{align}
That is, higher-order mutual information resolves strictly fewer logical gates by value and~none by sign. In~fact, the~higher-order mutual information of a logic gate can never be positive, because it is bounded from above by the minimum of the pairwise mutual information, which is always zero for the pair of inputs. Because~all entropy-based quantities inherit the degeneracy summarised in Table~\ref{tab:logicGateEntropies}, neither the mutual information nor its dual can increase the resolving power (see Table~\ref{tab:logicGateInOuteractions}).

\begin{table}[H]
\caption{The marginal entropies of variables in a logic gate are degenerate in XOR$\sim$XNOR and AND$\sim$OR$\sim$NAND$\sim$NOR.\label{tab:logicGateEntropies}}

\setlength{\tabcolsep}{7.5mm}  

\resizebox{\textwidth}{!}{
	\begin{tabular}{lrrrrc}
	\toprule
	$\boldsymbol{\mathcal{G}}$ & $\substack{\textbf{\emph{H(A)}}\\\textbf{\emph{=H(B)}}}$ & $\emph{\textbf{H(C)}}$ &  $\emph{\textbf{H(A,B)}}$ & $\substack{\textbf{\emph{H(A,C)}}\\\emph{\textbf{=H(B, C)}}}$ &    $\emph{\textbf{H(A,B,C)}}$ \\
	\midrule
	
	XNOR & $1$ & $1$ 	& $2$ & $2$	& $2$\\
	\rule{0pt}{3ex}XOR 	& $1$ & $1$ 	& $2$ & $2$	& $2$\\
	\rule{0pt}{3ex}AND 	& $1$ & $\log \frac{3^{3/4}}{4}$ & $2$ & $\frac{3}{2}$	& $2$\\
	\rule{0pt}{3ex}OR 	& $1$ & $\log \frac{3^{3/4}}{4}$ & $2$ & $\frac{3}{2}$	& $2$\\
	\rule{0pt}{3ex}NAND & $1$ & $\log \frac{3^{3/4}}{4}$ & $2$ & $\frac{3}{2}$	& $2$\\
	\rule{0pt}{3ex}NOR 	& $1$ & $\log \frac{3^{3/4}}{4}$ & $2$ & $\frac{3}{2}$	& $2$\\
	\bottomrule
	\end{tabular}
}
\end{table}

The logic gate interactions and their duals are summarised in Table~\ref{tab:logicGateInOuteractions}, where it can be seen that neither $I^{*\mathcal{G}}_{C} = I^{\mathcal{G}}_{ABC} + I^{\mathcal{G}}_{AB}$ nor $I^{*\mathcal{G}}_{A}$ improve the resolution beyond that of the 3-point interaction. However, the~3-point interaction requires $2^3=8$ probabilities to achieve this resolving power, whereas $I^{*\mathcal{G}}_{C} = \frac{p_{111} p_{001}}{p_{101}p_{011}}$ achieves the same resolving power with just four~probabilities.

However, note that because of a difference in sign convention dual mutual information is a difference between two mutual information quantities, while dual interactions are a sum of two interactions. Based on this, we can consider the difference of two interactions and define a new quantity $J^{*\mathcal{G}}_{A} = I^{\mathcal{G}}_{ABC} - I^{\mathcal{G}}_{BC}$. We refer to this as a $J$-interaction. When the MFIs are interpreted in the context of an energy-based model, such as an Ising model or a restricted Boltzmann machine, then the interactions have dimensions of energy, meaning that the $J$-interactions correspond to the difference in the energy contribution between a triplet and a pair. These $J$-interactions of the input nodes $A$ and $B$ assign a different value to each logic gate $\mathcal{G}$, and~the symmetric $J$-interaction $\overline{J}^{*\mathcal{G}} = J^{*\mathcal{G}}_A J^{*\mathcal{G}}_B J^{*\mathcal{G}}_C$, analogous to the symmetric deltas from \citep{Galas2014}, inherits the perfect resolution from $J^{*\mathcal{G}}_{A}$. 

Note that while $J^{*\mathcal{G}}_A = J^{*\mathcal{G}}_B$ both have perfect resolution, $J^{*\mathcal{G}}_C = I^{\mathcal{G}}_{ABC} - I^{\mathcal{G}}_{AB}$ does not improve the resolution beyond that of the 3-point interaction. This results from the fact that in logic gates we have $I^{\mathcal{G}}_{ABC} = -2 I^{\mathcal{G}}_{AB}$, meaning that $I^{\mathcal{G}}_{ABC}$ and $I^{\mathcal{G}}_{AB}$ contain the same information. To~see this, note that
\begin{align}
	I^{\mathcal{G}}_{ABC} + 2 I^{\mathcal{G}}_{AB} &= \log \frac{p_{111} p_{001} p_{110} p_{000}}{p_{101} p_{011} p_{010} p_{100}}
	\intertext{Because the logic gates are symmetric in their inputs, \ie $\forall i, j ~ p_{ijk} = p_{jik}$, this can be rewritten~as} 
  I^{\mathcal{G}}_{ABC} + 2 I^{\mathcal{G}}_{AB} &= \log \frac{(p_{111} p_{110}) (p_{001} p_{000})}{(p_{101} p_{100}) (p_{101} p_{100})}
	\intertext{Each of these terms in brackets has the form $(p_{ij1} p_{ij0})$. Because these are two contradicting states, this product reduces to $\epsilon p$ regardless of the truth table of $\mathcal{G}$:}
	&= \log \frac{\epsilon^2 p^2}{\epsilon^2 p^2} = 0
\end{align}
Note that this pattern could already be observed in~Table~\ref{tab:logicGateInOuteractions}, though~it was not yet explained.

Thus, the $J$-interactions of the input nodes uniquely assign a value to each gate proportional to the synergy of its logic. The~hierarchy is $J^{*\text{XNOR}}_{A} > J^{*\text{NOR}}_{A} > J^{*\text{AND}}_{A}$, which is mirrored for the respective logical complements. XNOR is indeed the most synergistic, while~NOR is more synergistic than AND with respect to observing a 0 in one of the inputs; in a NOR gate, a~0 in the input provides no information on the output, while it completely fixes the output of an AND gate. Because the interactions are defined in a context of 0s, they order the synergy~accordingly.

\begin{table}[H]
\caption{While the interactions leave certain gates indistinguishable, the~dual $J$-interactions of the inputs are unique to each gate. The~reported decimal values are rounded to three digits; as~before, $I = 4 \log \frac{p}{\epsilon}$.  \label{tab:logicGateInOuteractions}}

\begin{adjustwidth}{-\extralength}{0cm}
\centering 

\setlength{\tabcolsep}{4.5mm}  

\resizebox{\fulllength}{!}{
		\begin{tabular}{lrrrrrrrrrrr}
		\toprule
		 $\boldsymbol{\mathcal{G}}$ & \boldmath{$MI_{ABC}$} & \boldmath{$MI_{BC}$} & \boldmath{$MI^*_{A}$} &\boldmath{$I^\mathcal{G}_{ABC}$}  & \boldmath{$I^\mathcal{G}_{AB}$} &\boldmath{$I^\mathcal{G}_{BC}$} &\boldmath{$I^{*\mathcal{G}}_{A}$} &\boldmath{$I^{*\mathcal{G}}_{C}$} &\boldmath{$J^{*\mathcal{G}}_{A}$} &\boldmath{$J^{*\mathcal{G}}_{C}$} & \boldmath{$\overline{J}^{*\mathcal{G}}$}\\
		\midrule
		XNOR & $-1 $     & 0     & $-1$	 			& $I$               & $-\frac{1}{2}I$  &  $-\frac{1}{2}I$ 	& $\frac{1}{2}I$   & $\frac{1}{2}I$    & $\frac{3}{2}I$    & $\frac{3}{2}I$    & $\frac{27}{8}I^3$  \\
		\rule{0pt}{3ex}XOR  & $-1 $     & 0     & $-1$ 				& $-I$              & $\frac{1}{2}I$   &  $\frac{1}{2}I$  	& $-\frac{1}{2}I$  & $-\frac{1}{2}I$   & $-\frac{3}{2}I$   & $-\frac{3}{2}I$   & $-\frac{27}{8}I^3$ \\
		\rule{0pt}{3ex}AND  & $-0.189 $ & 0.311 & $-\frac{1}{2}$ 	& $\frac{1}{2}I$    & $-\frac{1}{4}I$  &  $0$  				& $\frac{1}{2}I$   & $\frac{1}{4}I$    & $\frac{1}{2}I$    & $\frac{3}{4}I$    & $\frac{3}{16}I^3$ \\
		\rule{0pt}{3ex}OR   & $-0.189 $ & 0.311 & $-\frac{1}{2}$ 	& $-\frac{1}{2}I$   & $\frac{1}{4}I$   &  $\frac{1}{2}I$  	& $0$              & $-\frac{1}{4}I$   & $-I$              & $-\frac{3}{4}I$   & $-\frac{3}{4}I^3$ \\
		\rule{0pt}{3ex}NAND & $-0.189 $ & 0.311 & $-\frac{1}{2}$ 	& $-\frac{1}{2}I$   & $\frac{1}{4}I$   &  $0$   			& $-\frac{1}{2}I$  & $-\frac{1}{4}I$   & $-\frac{1}{2}I$   & $-\frac{3}{4}I$   & $-\frac{3}{16}I^3$ \\
		\rule{0pt}{3ex}NOR  & $-0.189 $ & 0.311 & $-\frac{1}{2}$ 	& $\frac{1}{2}I$    & $-\frac{1}{4}I$  &  $-\frac{1}{2}I$  	& $0$              & $\frac{1}{4}I$    & $I$               & $\frac{3}{4}I$    & $\frac{3}{4}I^3$  \\
		\bottomrule
		\end{tabular}
		}
\end{adjustwidth}
	
	\end{table}

\FloatBarrier

\subsection{Interactions Distinguish Dynamics and Causal~Structures}

To illustrate how different association metrics reflect the underlying causal dynamics, consider data generated from a selection of three-node causal DAGs as follows. On a given DAG $\mathcal{G}$, first denote the set of nodes without parents, the~orphan nodes, by~$S_0$. Each orphan node in $S_0$ receives a random value drawn from a Bernoulli distribution, i.e.,~$P(X=1)=p$ and $P(X=0)=1-p$. Next, denote the set of children of orphan nodes as $S_1$. Each node in $S_1$ is then set to either the product of its parent nodes (for \textit{multiplicative} dynamics) or~the mean of its parent nodes (for \textit{additive} dynamics), plus some zero-mean Gaussian noise with variance $\sigma^2$. Note that for the fork and the chain this simply amounts to a noisy copying operation. All nodes are then rounded to a 0 or 1. A~set $S_2$ is then defined as the set of all children of nodes in $S_1$, and~these receive values using the same dynamics as before. As~long as the causal structure is acyclic, this algorithm terminates on a set of nodes $S_i$ that has no children. For~example, the~chain graph $A\to B \to C$ has $S_0=\{A\}$, $S_1=\{B\}$, $S_2=\{C\}$, and~$S_3=\emptyset$, at~which point the updating~terminates. 

\begin{figure}[H]

\includegraphics[width=13.5 cm]{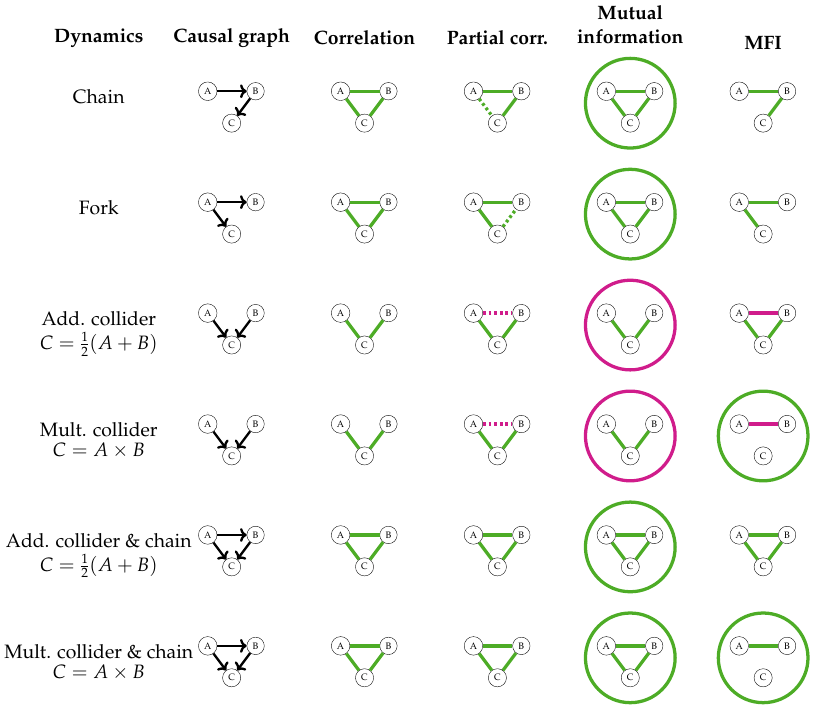}
	
	\caption{{Different}  
 causal dynamics lead to different association metrics. Green edges denote positive values, red edges denote negative values, circles denote a three-point quantity, and~dashed lines show edges with marginal significance (depending on $\sigma^2$). Correlations and mutual information cannot distinguish between most dynamics, and~while partial correlation can identify the correct pairwise relationships for~certain noise levels, it falls short of distinguishing additive from multiplicative dynamics. Only MFIs can distinguish between all six scenarios and~reveal the combinatorial effect of the multiplicative dynamics as a 3-point interaction. See Appendix \ref{A:DAGs} for the simulation parameters and raw numbers. This figure is reproduced with permission from the author of~\cite{myThesis}.}
	\label{fig:DAGs}
\end{figure}

Figure~\ref{fig:DAGs} shows the results for four different DAGs with~multiplicative and additive dynamics (though these are the same for forks and chains). The~six different dynamics are represented in four different DAGs, two different (Pearson) correlations, four different partial correlations, and~two different mutual information structures, which means that each of these descriptions is degenerate in some of the dynamics. The~shown pairwise partial correlations are the correlations among the residuals after a linear regression against the third variable. Because this is similar to conditioning on the third variable, it is somewhat analogous to the MFIs; in fact, when the variables are multivariate normal the partial correlations are encoded in the inverse covariance matrix and are equivalent to pairwise Ising interactions~\cite{Nguyen2017}. Indeed, it can be seen that the partial correlations are somewhat able to disentangle direct effects from indirect effects, although they~fail to distinguish additive from multiplicative dynamics. Note that only the sign of the association and its significance are represented, as the precise value depends on the noise level $\sigma^2$. The~rightmost column shows that the MFIs assign a unique association structure to each of the dynamics, distinguish between direct and indirect effects, and~reveal multiplicative dynamics as a 3-point interaction while identifying additive dynamics as a purely pairwise process. Finally, note that both the partial correlation and the MFIs assign a negative association to the parent nodes in a collider structure. This reflects that two nodes become dependent when conditioned on a common effect (cf. Berkson's paradox), a~phenomenon already found in partial correlations of metabolomic data in \citep{GGM}. The~mutual information is affected by Berkson's paradox as well, revealed through the negative three-point mutual information. This negative three-point is a direct effect from conditioning on the common effect $C$, as on colliders $MI(A, B, C) = MI(A, B) - MI(A, B \mid C) = -MI(A, B \mid C)$, because~the mutual information among the independent inputs $A$ and $B$ vanishes by~definition.

\FloatBarrier

\subsection{Higher-Order Categorical Interactions Distinguish Dyadic and Triadic~Distributions}

That the interactions have such resolving power over distributions of binary variables is perhaps not very surprising in light of the universality of RBMs with respect to this class of distributions. More surprisingly, their resolving power extends to the case of categorical variables. In~\citep{dyTriadic}, the~authors introduced two distributions, the~dyadic and triadic distributions, which are indistinguishable by almost all commonly used information measures (\ie Shannon, Renyi(2), residual, and~Tsallis entropy, co-information, total correlation, CAEKL mutual information, interaction information, Wyner, exact, functional, and~MSS common information, perplexity, disequilibrium, and~LMRP and TSE complexities).

The two distributions are defined on three variables, each taking a value in a four-letter alphabet $\{0, 1, 2, 3\}$. The~joint probabilities are summarised in Table~\ref{tab:dyTriadic}. To~construct the distributions, each category is represented as a binary string ($\{0,1,2,3\} \to \{00, 01, 10, 11\}$), leading to new variables $\{X_0, X_1, Y_0, Y_1, Z_0, Z_1\}$. The~dyadic distribution is constructed by linking these new variables with pairwise rules $X_0 = Y_1, Y_0 = Z_1, Z_0 = X_1$, while the triadic distribution is constructed with triplet rules $X_0 + Y_0 + Z_0 = 0 \mod 2$ and~$X_1=Y_1=Z_1$. The~resulting binary strings are then reinterpreted as categorical variables to produce Table~\ref{tab:dyTriadic}. 

The authors of \citep{dyTriadic} found that no Shannon-like measure can distinguish between the two distributions, and argued that the partial information decomposition, which is different for the two distributions, is not a natural information measure, as it has to single out one of the variables as an output. To~calculate model-free categorical interactions between the variables, we can set the probabilities of the states in Table~\ref{tab:dyTriadic} uniformly to $p=(1-(64-8)\epsilon)/8$ and~those of the other states to $\epsilon$ (\ie a normalised uniform distribution over legal states). There are a total of $6^3=216$ interactions such that $x_1>x_0,~ y_1>y_0,~ z_1>z_0$. Each of these can be written as
\begin{align}
    \nonumber & I_{XYZ}(x_0 \to x_1; y_0 \to y_1; z_0 \to z_1)  =\\
     \nonumber \log &\frac{p\Big(X=x_1, Y=y_1, Z=z_1 \mid \underline{X}=0\Big)}{p\Big(X=x_0, Y=y_0,, Z=z_0\mid \underline{X}=0\Big)}  
        \frac{p\Big(X=x_1, Y=y_0, Z=z_0 \mid \underline{X}=0\Big)}{p\Big(X=x_0, Y=y_1,, Z=z_1\mid \underline{X}=0\Big)}\\
        \times &\frac{p\Big(X=x_0, Y=y_1, Z=z_0 \mid \underline{X}=0\Big)}{p\Big(X=x_1, Y=y_0,, Z=z_1\mid \underline{X}=0\Big)}
        \frac{p\Big(X=x_0, Y=y_0, Z=z_1 \mid \underline{X}=0\Big)}{p\Big(X=x_1, Y=y_1,, Z=z_0\mid \underline{X}=0\Big)}
\end{align}

\vspace{-18pt}

Of particular interest here are the two quantities $I_{XYZ}(0 \to 3; 0 \to 3; 0 \to 3)$ and~$\overline{I}_{XYZ} = \sum_{x_0, x_1, y_0, y_1, z_0, z_1} I_{XYZ}(x_0 \to x_1; y_0 \to y_1; z_0 \to z_1)$, where the sum is over all values such that $x_1>x_0, y_1>y_0, z_1>z_0$, as all possible pairs necessarily sum to zero because $I_{XYZ}(x_0 \to x_1; y_0 \to y_1; z_0 \to z_1) = - I_{XYZ}(x_1 \to x_0; y_0 \to y_1; z_0 \to z_1) $. For~the dyadic distribution, we have\vspace{-6pt}
\begin{align}
	I^\text{Dy}_{XYZ}(0 \to 3; 0 \to 3; 0 \to 3) = \log \frac{p \epsilon^3}{p \epsilon^3}=0,\\
	\intertext{while for the triadic distribution we have}
	I^\text{Tri}_{XYZ}(0 \to 3; 0 \to 3; 0 \to 3) = \log \frac{\epsilon^4}{p \epsilon^3} = \log\frac{\epsilon}{p}
\end{align}
Thus, this particular 3-point interaction is zero for the dyadic distribution and~negative for the triadic distribution. The~sum over all three points (see Appendix \ref{A:pythonCode} for details) is provided by
\begin{align}
	\overline{I}^\text{Dy}_{XYZ} &= \log 1 = 0\\
	\overline{I}^\text{Tri}_{XYZ} &= 64 \log \frac{\epsilon}{p}
\end{align}

That is, the~additively symmetrised 3-point interaction is zero for the dyadic distribution and~strongly negative for the triadic distribution. These two distributions, which are indistinguishable in terms of their information structure, are distinguishable by their model-free interactions, which accurately reflect the higher-order nature of the triadic~distribution.

\begin{table}[H]
\caption{The joint probability of the dyadic and triadic distributions~\cite{dyTriadic}. All other states have a probability of~zero.}
\begin{tabularx}{\textwidth}{CCCC}
\toprule
\multicolumn{4}{c}{\textbf{Dyadic}}\\
\midrule
\textbf{X} & \textbf{Y} & \textbf{Z} & \textbf{P}\\
\midrule 
0 & 0 & 0 & 1 / 8\\
0 & 2 & 1 & 1 / 8 \\
1 & 0 & 2 & 1 / 8 \\
1 & 2 & 3 & 1 / 8 \\
2 & 1 & 0 & 1 / 8 \\
2 & 3 & 1 & 1 / 8 \\
3 & 1 & 2 & 1 / 8 \\
3 & 3 & 3 & 1 / 8 \\
\midrule
\multicolumn{4}{c}{\textbf{Triadic}}\\
\midrule
\textbf{X} & \textbf{Y} & \textbf{Z} & \textbf{P}\\
\midrule
0 & 0 & 0 & 1 / 8 \\
1 & 1 & 1 & 1 / 8 \\
0 & 2 & 2 & 1 / 8 \\
1 & 3 & 3 & 1 / 8 \\
2 & 0 & 2 & 1 / 8 \\
3 & 1 & 3 & 1 / 8 \\
2 & 2 & 0 & 1 / 8 \\
3 & 3 & 1 & 1 / 8 \\
\bottomrule
\end{tabularx}   

\label{tab:dyTriadic}
\end{table}

\FloatBarrier

\section{Discussion \label{sec:discussion}}

In this paper, we have related the model-free interactions introduced in \citep{AvaSjoerd} to information theory by defining them as Möbius inversions of surprisal on the same lattice that relates mutual information to entropy. We then invert the order of the lattice and compute the order-dual to the mutual information, which turns out to be a generalisation of differential mutual information. Similarly, the~order-dual of interaction turns out to be interaction in a different context. Both the interactions and the dual interactions are able to distinguish all six logic gates by value and sign. Moreover, their absolute strength reflects the synergy within the logic gate. In~simulations, the~interactions were able to perfectly distinguish six kinds of causal dynamics that are partially indistinguishable to Pearson/partial correlations, causal graphs, and~mutual information. Finally, we considered dyadic and triadic distributions constructed using pairwise and~higher-order rules, respectively. While these two distributions are indistinguishable in terms of their Shannon information, they~have different categorical MFIs that reflect the order of the construction~rules. 

One might wonder why the interactions enjoy this advantage over entropy-based quantities. The~most obvious difference is that the interactions are defined in a pointwise way, \ie in terms of the surprisal of particular states, whereas entropy is the expected surprisal across an ensemble of states. Furthermore, the~MFIs can be interpreted as interactions in an Ising model and as effective couplings in a restricted Boltzmann machine. As both these models are known to be universal approximators with respect to positive discrete probability distributions, the~MFIs should be able to characterise all such distributions. What is not immediately obvious is that the kinds of interactions that characterise a distribution should reflect properties of that distribution, such as the difference between direct and indirect effects and~the presence of higher-order structure. However, in~the various examples covered in this manuscript the~interactions turn out to intuitively align with properties of the process used to generate the data. While the~stringent conditioning on variables not considered in the interaction might make it tempting to interpret an MFI as a causal or interventional quantity, it is important to be very careful when doing this. Assigning a causal interpretation to statistical inferences, whether in Pearl's graphical do-calculus \citep{Pearl2000} or in Rubin's potential outcomes framework \citep{Imbens2015}, requires further (often untestable) assumptions and analysis of the system in order to determine whether a causal effect is identifiable and~which variables to control for. In~contrast, an~MFI is simply defined by conditioning on all observed variables, makes no reference to interventions or counterfactuals, and~does not specify a direction of the effect. While in a controlled and simple setting the~MFIs can be expressed in terms of causal average treatment effects \citep{AvaSjoerd}, a~causal interpretation is not justifiable in~general. 

Moreover, the~stringency in the conditioning might worry the attentive reader. Estimating $\log p(X=1, Y=1, T=0)$ directly from data means counting states such as $(X, Y, T_1, T_2, \ldots , T_N) = (1, 1, 0, 0, \ldots 0)$, which for sufficiently large $N$ are rare in most datasets. Appendix \ref{A:MBs} shows how to use the causal graph to construct Markov blankets, making such estimation tractable when full conditioning is too stringent. In~an upcoming paper, we address this issue by estimating the graph of conditional dependencies, allowing for~successful calculation of MFIs up to the fifth order in gene expression~data.

One major limitation of MFIs is that they are only defined on binary or categorical variables, whereas many other association metrics are defined for ordinal and continuous variables as well. As states of continuous variables no longer form a lattice, it is hard to see how the definition of MFIs could be extended to include these~cases.

Finally, it is worth noting that the structure of different lattices has guided much of this research. That Boolean algebras are important in defining higher-order structure is not surprising, as they are the stage on which the inclusion--exclusion principle can be generalised \citep{enumComb}. However, it is not only their order-reversed duals that lead to meaningful definitions; completely unrelated lattices do as well. For~example, the~Möbius inversion on the lattice of ordinal variables from Figure~\ref{fig:lattice_cats} and the redundancy lattices in the partial information decomposition \citep{PID} both lead to new and sensible definitions of information-theoretic quantities. Furthermore, the~notion of Möbius inversion has been generalised to a more general class of categories~\cite{Leinster2012}, of~which posets are a special case. A~systematic investigation of information-theoretic quantities in this richer context would be most~interesting.

\vspace{6pt}

\funding{This research was funded by Medical Research Council grant number MR/N013166/1.}

\dataavailability{No new data were created or analyzed in this study. Data sharing is not applicable to this article.} 

\acknowledgments{The author is grateful for the insightful discussions of model-free interactions, causality, and~RBMs with Ava Khamseh, Sjoerd Beentjes, Chris Ponting, and~Luigi Del Debbio. The~author also thanks John Baez for a helpful conversation on the role of surprisal in information theory. The~author further thanks Sjoerd Beentjes for reading and commenting on an early version of this work, and~the reviewers for their helpful suggestions. A.J. is supported by an MRC Precision Medicine Grant (MR/N013166/1).}

\conflictsofinterest{The funders had no role in the design of the study, in the collection, analysis, or~interpretation of data, in the writing of the manuscript, or in the decision to publish the~results.}



\abbreviations{Abbreviations}{
The following abbreviations are used in this manuscript:\\

\noindent 
\begin{tabular}{@{}ll}
MI & Mutual Information\\
MFI & Model-Free Interaction\\
DAG & Directed Acyclic Graph\\
MB & Markov Blanket\\
PID & Partial Information d=Decomposition\\
i.i.d. & independent and identically distributed
\end{tabular}
}


\appendixtitles{no} 
\appendixstart
\appendix
\section[\appendixname~\thesection]{}
\appendixtitles{yes}
\subsection[\appendixname~\thesubsection]{Markov Blankets} \label{A:MBs}
Estimating the interaction in Definition \ref{def:MFIs} from data involves estimating the probabilities of certain states occurring. While we do not have access to the true probabilities, we~can rewrite the interactions in terms of expectation values. Note that all interactions involve factors of the type
\begin{align}
  \frac{p(X=1, Y=y \mid Z=0)}{p(X=0, Y=y \mid Z=0)} &= \frac{p(X=1 \mid Y=y, Z=0)}{p(X=0\mid Y=y, Z=0)}\\
  &= \frac{p(X=1 \mid Y=y, Z=0)}{1-p(X=1 \mid Y=y, Z=0)}\\
  &= \frac{\mathbb{E}[X\mid Y=y, Z=0]}{1-\mathbb{E}[X\mid Y=y, Z=0]}
  \intertext{because}
  \mathbb{E}[X \mid Z=z] = \sum_{x \in \{0, 1\}} p(X=x\mid Z=z) ~ x &= p(X=1\mid Z=z)
\end{align}
This allows us to write the 2-point interaction, \eg as follows:
\begin{align}
    I_{ij} &= \log \frac{\mathbb{E}\left(X_{i} | X_{j}=1, \underline{X}=0\right)}{\mathbb{E}\left(X_{i} | X_{j}=0, \underline{X}=0\right)} \frac{\left(1-\mathbb{E}\left(X_{i} | X_{j}=0, \underline{X}=0\right)\right)}{\left(1-\mathbb{E}\left(X_{i} | X_{j}=1, \underline{X}=0\right)\right)} \label{TL_expect} 
\end{align}

Although expectation values are theoretical quantities, not empirical ones, sample means can be used as unbiased estimators to estimate each term in \eqref{TL_expect}. The~stringent conditioning in this estimator can make the number of samples that satisfy the conditioning very small, which results in the estimates having large variance on different finite samples. Note that if we can find a subset of variables $\text{MB}_{X_i}$ such that $X_i \indep X_k \mid \text{MB}_{X_i} ~~\forall  X_k \notin \text{MB}_{X_i}$ and $i \neq k$ (in causal language, a set of variables $\text{MB}_{X_i}$ that d-separates $X_i$ from the rest), then we only have to condition on $\text{MB}_{X_i}$ in \eqref{TL_expect}, reducing the variance of our estimator. Such a set $\text{MB}_{X_i}$ is called a \textit{Markov Blanket} of the node $X_i$. {There
 has recently been a certain degree of confusion around the notion of Markov blankets in biology, specifically with respect to their use in the free energy principle in neuroscience contexts. Here, a Markov blanket refers to the notion of a \textit{Pearl blanket} in the language of~\cite{Bruineberg2022}}. Because conditioning on fewer variables should reduce the variance of the estimate by increasing the number of samples that can be used for the estimation, we are generally interested in finding the smallest Markov blanket. This minimal Markov blanket is called the Markov~boundary.

Finding such minimal Markov blankets is hard; in fact, because it requires testing each possible conditional dependency between the variables, we claim here (without proof) that it is \textit{causal discovery}-hard, \ie if~such a graph~exists it is at least as computationally complex as constructing a causal DAG consistent with the joint probability distribution.

\subsection[\appendixname~\thesubsection]{Proofs} \label{A:proofs} 
Markov blankets are not only a computational trick; in theory, only variables that are in each other's Markov blanket can share a nonzero interaction. To~illustrate this, first note that the property of being in a variable's Markov blanket is symmetric:
\begin{Proposition}[symmetry of Markov blankets]\label{prop:symOfMBs} 
	Let $X$ be a set of variables with joint distribution $p(X)$ and let $A \in X$ and $B \in X$ such that $A \neq B$. We denote the minimal Markov blanket of $X$ by $\text{MB}_X$. Then, $A \in \text{MB}_B \iff B \in \text{MB}_A$, and~we can say that $A$ and $B$ are Markov-connected. 
\end{Proposition}
\begin{proof}
Let $Y = X \setminus \{A, B\}$. Then,
\begin{align}
  A \not\in \text{MB}_B \implies p(B \mid A, Y) &= p(B \mid Y)
  \intertext{Consider that}
  p(A \mid B, Y) &= \frac{p(A, B \mid Y)}{p(B\mid Y)}\\
   &= \frac{p(B\mid A, Y) p(A, \mid Y)}{p(B \mid Y)}\\
   &= p(A\mid Y)
\end{align}
which means that $B \not\in \text{MB}_A$. Because $A \not\in \text{MB}_B \iff B \not\in \text{MB}_A$ holds, its negation holds as well, which completes the proof.
\end{proof}

This definition of Markov connectedness allows us to state the following. 

\begin{Theorem}[only Markov-connected variables can interact]\label{thm:markovConnected}
	A model-free n-point interaction $I_{1\ldots n}$ can only be nonzero when all variables $S = \{X_1, \ldots, X_n\}$ are mutually Markov-connected. 
\end{Theorem}
\begin{proof}
	Let $X$ be a set of variables with joint distribution $p(X)$, let $S = \{X_1, \ldots, X_n\}$, and~let $\underline{X} = X \setminus S$. Consider the definition of an $n$-point interaction among $S$:\vspace{-6pt}
\begin{align}
  I_{1\ldots n} &= \prod_{i=1}^n \frac{\partial}{\partial X_i} \log p(X_1, \ldots, X_n \mid \underline{X}=0)\\
  &= \left(\prod_{i=1}^{n-1} \frac{\partial}{\partial X_i}  \right) \frac{\partial}{\partial X_n}\log p(X_1, \ldots, X_n \mid \underline{X}=0)\\
  &=  \left(\prod_{i=1}^{n-1} \frac{\partial}{\partial X_i}  \right) \log \frac{p(X_n=1 \mid X_1, \ldots, X_{n-1}, \underline{X}=0)}{p(X_n=0 \mid X_1, \ldots, X_{n-1}, \underline{X}=0)}\\
  &=  \left(\prod_{i=1}^{n-1} \frac{\partial}{\partial X_i}  \right) \log \frac{p(X_n=1 \mid S\setminus X_n, \underline{X}=0)}{p(X_n=0 \mid S\setminus X_n, \underline{X}=0)}
  \intertext{Now, if~$\exists X_j \in S$ such that $X_j \not\in \text{MB}_{X_n}$, we do not need to condition on $X_j$ and can write this as}
  I_{1\ldots n} &=  \left(\prod_{i=1}^{n-1} \frac{\partial}{\partial X_i}  \right) \log \frac{p(X_n=1 \mid S\setminus \{X_j, X_n\}, \underline{X}=0)}{p(X_n=0 \mid S\setminus \{X_j, X_n\}, \underline{X}=0)}\\
  &=  \left(\prod_{\substack{i=1\\i\neq j}}^{n-1} \frac{\partial}{\partial X_i}  \right) \left(\frac{\partial}{\partial X_j} \log \frac{p(X_n=1 \mid S\setminus \{X_j, X_n\}, \underline{X}=0)}{p(X_n=0 \mid S\setminus \{X_j, X_n\}, \underline{X}=0)}\right)\\
  &=0
\end{align}
as the probabilities no longer involve $X_j$. Because $X_j$ was chosen arbitrarily, this must hold for all variables in $S$, which means that if any variable in $S$ is not in the Markov blanket of $X_n$ then the interaction $I_S$ vanishes:
\begin{align}
  S\setminus X_n \not\subset \text{MB}_{X_n} \implies I_S = 0
\end{align}
Furthermore, as the~indexing we chose for our variables was arbitrary, this must hold for any re-indexing, which means that
\begin{align}
  \forall X_i \in S:~~ S\setminus X_i \not\subset \text{MB}_{X_i} \implies I_S = 0
\end{align}
This in turn means that all variables in $S$ must be Markov-connected in order for the interaction $I_S$ to be nonzero. 
\end{proof}

Thus, knowledge of the causal graph aids estimation in two ways: it shrinks the variance of the estimates by relaxing the conditioning, and~it identifies the interactions that could be~nonzero.\\

When knowledge of the causal graph is imperfect, it is possible to accidentally exclude a variable from a Markov blanket and thereby undercondition the relevant probabilities. The~resulting error can be expressed in terms of the mutual information between the variables, as follows.
\begin{Proposition}[underconditioning bias]
Let $S$ be a set of random variables with probability distribution $p(S)$, let $X, Y$, and~let $Z$ be three disjoint subsets of $S$. Then, omitting $Y$ from the conditioning set results in a bias determined by (and~linear in) the~pointwise mutual information that $Y=0$ provides about the states of $X$:\vspace{-6pt}
\begin{align}
  I_{X\mid Y Z} - I_{X\mid Z} = \left(\prod_{i=1}^{|X|} \frac{\partial}{\partial x_i}\right)\text{pmi}(X=x, Y=0 \mid Z=0)
\end{align}
\end{Proposition}
\begin{proof}

The pointwise mutual information (pmi) is defined as
\begin{align}
	\text{pmi}(X=x, Y=y) &= \log \frac{p(X=x, Y=y)}{p(X=x)p(Y=y)}
	\intertext{Note that}
	p(X=x_1\mid Y=y, Z=z) &= \frac{p(X=x_1, Y=y \mid Z=z)}{p(Y=y\mid Z=z)}
	\intertext{meaning that we can write}
	p(X=x_1\mid Y=y, Z=z) &= e^{\text{pmi}(X=x_1, Y=y \mid Z=z)} p(X=x_1\mid Z=z)
\end{align}
That is, not conditioning on $Y=y$ results in an error in the estimate of ${p(X=x_1\mid Y=y,}$ ${Z=z)}$ that is exponential in the $Z$-conditional pmi of $X$ and $Y$. However, consider the interaction among $X$,
\begin{align}
  I_X = I_{X\mid Y Z} &= \left(\prod_{i=1}^{|X|} \frac{\partial}{\partial x_i}\right) \log p(X=x \mid Y=0, Z=0)\\
  &= \left(\prod_{i=1}^{|X|} \frac{\partial}{\partial x_i}\right) \left(\log p(X=x \mid  Z=0) + \text{pmi}(X=x, Y=0 \mid Z=0)\right)\\
  &= I_{X\mid Z} + \left(\prod_{i=1}^{|X|} \frac{\partial}{\partial x_i}\right) \text{pmi}(X=x, Y=0 \mid Z=0)
\end{align}

That is, the~error in the interaction as a result of not conditioning on the right variables is linear in terms of the difference between the pmi values of different states.
\end{proof}

\subsection[\appendixname~\thesubsection]{Numerics of Causal Structures}\label{A:DAGs}

Tables~\ref{tab:firstDAG}--\ref{tab:lastDAG} are taken from~\cite{myThesis} with permission from the author, and~list the precise values leading to Figure~\ref{fig:DAGs}. From~each graph, 100k samples were generated using $p=0.5$ and $\sigma=0.4$. To~quantify the significance value of the interactions, the~data were bootstrap resampled 1k times, resulting in the definition of $F$ as the fraction of resampled interactions having a different sign from the original interaction. The~smaller $F$ is, the~more significant the interaction.

\begin{table}[H]
%

\caption{Chain. \label{tab:firstDAG}}

\includegraphics[width=2.5 cm]{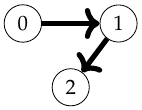}

{\tablesize{\footnotesize}

	\begin{tabularx}{\textwidth}{llrrrrrrr}
		\toprule
		{} &      \textbf{Genes} &  \textbf{Interaction} &      \textbf{F} &  \textbf{Pearson cor.} &  \textbf{Pearson cor. p} &  \textbf{Partial cor.} &  \textbf{Partial cor. p} &     \textbf{MI} \\
		\midrule
		0 &     [0, 1] &        4.281 &  0.000 &         0.790 &             0.0 &         0.635 &       0.000 
 $\times$ 10$^{+0}$ &  0.515 \\
		1 &     [0, 2] &        0.056 &  0.117 &         0.622 &             0.0 &         0.031 &       2.261 $\times$ 10$^{-23}$ &  0.301 \\
		2 &     [1, 2] &        4.249 &  0.000 &         0.786 &             0.0 &         0.628 &       0.000 $\times$ 10$^{+0}$ &  0.510 \\
		3 &  [0, 1, 2] &       $-$0.052 &  0.217 &           NaN &             NaN &           NaN &             NaN &  0.300 \\
		\bottomrule
		\end{tabularx}}

\end{table}
\unskip

\begin{table}[H]
%

	\caption{Fork.}
\includegraphics[width=2.5 cm]{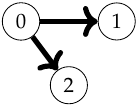}

	\begin{tabularx}{\textwidth}{llrrrrrrr}
		\toprule
		{} &      \textbf{Genes} &  \textbf{Interaction} &      \textbf{F} &  \textbf{Pearson cor.} &  \textbf{Pearson cor. p} &  \textbf{Partial cor.} &  \textbf{Partial cor. p} &     \textbf{MI} \\
		\midrule
		0 &     [0, 1] &        4.268 &  0.000 &         0.789 &             0.0 &         0.634 &       0.000 $\times$ 10$^{+0}$ &  0.514 \\
		1 &     [0, 2] &        4.257 &  0.000 &         0.788 &             0.0 &         0.632 &       0.000 $\times$ 10$^{+0}$ &  0.512 \\
		2 &     [1, 2] &       $-$0.014 &  0.376 &         0.622 &             0.0 &         0.028 &       6.518 $\times$ 10$^{-19}$ &  0.300 \\
		3 &  [0, 1, 2] &        0.020 &  0.376 &           NaN &             NaN &           NaN &             NaN &  0.300 \\
		\bottomrule
		\end{tabularx}

\end{table}
\unskip

\begin{table}[H]
%

	\caption{Additive~collider.}
	
	\includegraphics[width=2.5 cm]{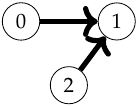}
	\tablesize{\scriptsize}

	\begin{tabularx}{\textwidth}{Xlrrrrrrr}
		\toprule
		{} &      \textbf{Genes} &  \textbf{Interaction} &      \textbf{F} &  \textbf{Pearson cor.} &  \textbf{Pearson cor. p} &  \textbf{Partial cor.} &  \textbf{Partial cor. p} &     \textbf{MI} \\
		\midrule
		0 &     [0, 1] &        2.144 &  0.000 &         0.395 &           0.000 &         0.505 &       0.000 $\times$ 10$^{+0}$ &  1.154 $\times$ 10$^{-1}$ \\
		1 &     [0, 2] &       $-$0.989 &  0.000 &        $-$0.002 &           0.593 &        $-$0.070 &      5.172 $\times$ 10$^{-109}$ &  2.059 $\times$ 10$^{-6}$ \\
		2 &     [1, 2] &        2.144 &  0.000 &         0.395 &           0.000 &         0.505 &       0.000 $\times$ 10$^{+0}$ &  1.154 $\times$ 10$^{-1}$ \\
		3 &  [0, 1, 2] &        0.003 &  0.438 &           NaN &             NaN &           NaN &             NaN & $-$2.678 $\times$ 10$^{-2}$ \\
		\bottomrule
		\end{tabularx}

\end{table}
\unskip

\begin{table}[H]
%

	\caption{Multiplicative~collider.}
	
		\includegraphics[width=2.5 cm]{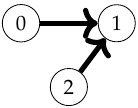}

	\begin{tabularx}{\textwidth}{Xlrrrrrrr}
		\toprule
		{} &      \textbf{Genes} &  \textbf{Interaction} &      \textbf{F} &  \textbf{Pearson cor.} &  \textbf{Pearson cor. p} &  \textbf{Partial cor.} &  \textbf{Partial cor. p} &     \textbf{MI} \\
		\midrule
		0 &     [0, 1] &        0.032 &  0.140 &         0.427 &           0.000 &         0.478 &       0.000 $\times$ 10$^{+0}$ &  1.403 $\times$ 10$^{-1}$ \\
		1 &     [0, 2] &       $-$2.156 &  0.000 &        $-$0.005 &           0.145 &        $-$40.087 &      1.463 $\times$ 10$^{-166}$ &  1.529 $\times$ 10$^{-5}$ \\
		2 &     [1, 2] &        0.036 &  0.109 &         0.429 &           0.000 &         0.480 &       0.000 $\times$ 10$^{+0}$ &  1.415 $\times$ 10$^{-1}$ \\
		3 &  [0, 1, 2] &        4.237 &  0.000 &           NaN &             NaN &           NaN &             NaN & $-$1.150 $\times$ 10$^{-1}$ \\
		\bottomrule
		\end{tabularx}
%
\end{table}
\unskip

\begin{table}[H]
%

\caption{Additive collider + chain.}
		\includegraphics[width=2.5 cm]{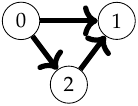}

	\tablesize{\footnotesize}
	\begin{tabularx}{\textwidth}{llrrrrrrr}
		\toprule
		{} &      \textbf{Genes} &  \textbf{Interaction} &      \textbf{F} &  \textbf{Pearson cor.} &  \textbf{Pearson cor. p} &  \textbf{Partial cor.} &  \textbf{Partial cor. p} &     \textbf{MI}  \\
		\midrule
		0 &     [0, 1] &        2.103 &  0.000 &         0.705 &             0.0 &         0.362 &             0.0 &  0.396 \\
		1 &     [0, 2] &        3.288 &  0.000 &         0.790 &             0.0 &         0.599 &             0.0 &  0.515 \\
		2 &     [1, 2] &        2.113 &  0.000 &         0.706 &             0.0 &         0.364 &             0.0 &  0.397 \\
		3 &  [0, 1, 2] &        0.050 &  0.162 &           NaN &             NaN &           NaN &             NaN &  0.335 \\
		\bottomrule
		\end{tabularx}
%
\end{table}
\unskip

\begin{table}[H]
%

\caption{Multiplicative collider + chain. \label{tab:lastDAG}}

		\includegraphics[width=2.5 cm]{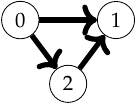}

	\begin{tabularx}{\textwidth}{llrrrrrrr}
		\toprule
		{} &      \textbf{Genes} &  \textbf{Interaction} &      \textbf{F} &  \textbf{Pearson cor.} &  \textbf{Pearson cor. p} &  \textbf{Partial cor.} &  \textbf{Partial cor. p} &     \textbf{MI} \\
		\midrule
		0 &     [0, 1] &       $-$0.017 &  0.342 &         0.709 &             0.0 &         0.365 &             0.0 &  0.403 \\
		1 &     [0, 2] &        2.094 &  0.000 &         0.786 &             0.0 &         0.596 &             0.0 &  0.510 \\
		2 &     [1, 2] &       $-$0.057 &  0.092 &         0.707 &             0.0 &         0.361 &             0.0 &  0.401 \\
		3 &  [0, 1, 2] &        4.359 &  0.000 &           NaN &             NaN &           NaN &             NaN &  0.293 \\
		\bottomrule
		\end{tabularx}
		
%
\end{table}

\FloatBarrier
\subsection[\appendixname~\thesubsection]{Python Code for Calculating Categorical Dyadic and Triadic Interactions\label{A:pythonCode}}

\begin{adjustwidth}{-\extralength}{0cm}
\centering 
\scriptsize 
{\begin{lstlisting}[language=Python]
dyadicStates = [['a', 'a', 'a'], ['a', 'c', 'b'], ['b', 'a', 'c'], ['b', 'c', 'd'], 
['c', 'b', 'a'], ['c', 'd', 'b'], ['d', 'b', 'c'], ['d', 'd', 'd']]

triadicStates = [['a', 'a', 'a'], ['a', 'c', 'c'], ['b', 'b', 'b'], ['b', 'd', 'd'], 
['c', 'a', 'c'], ['c', 'c', 'a'], ['d', 'b', 'd'], ['d', 'd', 'b']]

stateDict = {0: 'a', 1: 'b', 2:'c', 3: 'd'}

def catIntSymb(x0, x1, y0, y1, z0, z1, states):
    prob = lambda x, y, z: 'p' if [x, y, z] in states else~'e'
    
    num = prob(x1, y1, z1) + prob(x1, y0, z0) + prob(x0, y1, z0) + prob(x0, y0, z1)
    denom = prob(x1, y1, z0) + prob(x1, y0, z1) + prob(x0, y1, z1) + prob(x0, y0, z0)
    return (num, denom)

numDy = ''
denomDy = ''
numTri = ''
denomTri = ''

for x0 in range(4):
    for x1 in range(x0+1, 4):
        for y0 in range(4):
            for y1 in range(y0+1, 4):
                for z0 in range(4):
                    for z1 in range(z0+1, 4):
                        
                        nDy, dDy = catIntSymb(*[stateDict[x] for x in [x0, x1, y0, y1, z0, z1]], dyadicStates) 
                        numDy += nDy
                        denomDy += dDy
                        
                        nTri, dTri = catIntSymb(*[stateDict[x] for x in [x0, x1, y0, y1, z0, z1]], triadicStates) 
                        numTri += nTri
                        denomTri += dTri


print(f'Dyadic interaction: log (p^{numDy.count("p") - denomDy.count("p")} e^{numDy.count("e") - denomDy.count("e")})')
print(f'Triadic interaction: log (p^{numTri.count("p") - denomTri.count("p")} e^{numTri.count("e") - denomTri.count("e")})')

// Output: 

>> Dyadic interaction: log (p^0 e^0)
>> Triadic interaction: log (p^-64 e^64)
\end{lstlisting}}
\end{adjustwidth}

\FloatBarrier

\begin{adjustwidth}{-\extralength}{0cm}

\reftitle{References}


\PublishersNote{}
\end{adjustwidth}
\end{document}